\documentclass[12pt]{article}

\usepackage{amsmath}
\usepackage{amssymb}
\usepackage{mathrsfs}
\usepackage[T1]{fontenc}
\usepackage{amsthm}
\usepackage{enumerate}
\usepackage{fullpage}
\usepackage{multirow}
\usepackage{color}

\def\eqn{equation}
\def\cond{condition}
\def\tfn{transformation}

\def\dualn{dualization}
\def\soln{solution}
\def\fn{function}
\def\sm{sigma model}

\def\dd{Drinfel'd double}

\def\4diml{four-dimensional}
\def\-1{^{-1}}
\def\half{\frac{1}{2}}
\def\coor{coordinate}
\def\real{\mathbb{R}}

\def \unit {{\bf 1}}

\def\e{{\rm e}}
\def\cd{{\mathfrak d}}
\def\cg{{\mathfrak g}}

\def\wh{\widehat}

\def\sm{sigma model}
\def\pltp{Poisson--Lie T-pluralit}
\def\pltd{Poisson--Lie T-dualit}

\def\cf{{\mathcal {F}}}

\def\ghat{\hat g}
\newcommand{\prop}[2]{$\begin{array}{c}\text{#1},\\\text{#2}\\ \end{array}$}
\newcommand{\propp}[3]{$\begin{array}{c}\text{#1} \\ \text{#2},\\ \text{#3} \\ \end{array}$}

\newcommand{\diag}[4]{$\text{diag}\left(#1,#2,#3,#4\right)$}

\begin{document}

%\frontmatter

%%%%%%%%%%%%%%%%%%%%%%%%%%%%%%%%%%%%%%%%%%%%%%%%%%%%%%%%%%%%%%%%%%%%%%%%%%%%%%%
%% Titlepage
%%%%%%%%%%%%%%%%%%%%%%%%%%%%%%%%%%%%%%%%%%%%%%%%%%%%%%%%%%%%%%%%%%%%%%%%%%%%%%%

\title{Plane-parallel waves as duals of the flat background II: T-duality with spectators}
\author{Filip Petr\'asek, Ladislav Hlavat\'y\footnote{filip.petrasek@fjfi.cvut.cz, hlavaty@fjfi.cvut.cz}\\
{\em Faculty of Nuclear Sciences and Physical Engineering,} \\
Ivo Petr\footnote{ivo.petr@fit.cvut.cz}\\{\em Faculty of Information Technology} \\Czech Technical University in Prague}

\maketitle
\begin{abstract}

We give the classification of T-duals of the flat background in four
dimensions with respect to {one-, two-, and three-dimensional}
subgroups of the Poincar\'e group {using non-Abelian T-duality}
with spectators. As duals we find backgrounds for sigma models in
the form of plane-parallel waves or diagonalizable curved metrics
often with torsion. Among others, we find exactly solvable
time-dependent isotropic pp-wave, singular pp-waves, or generalized
{plane wave} (K-model).
%\keywords{Sigma Models; pp-wave background; String Duality.}

%\ccode{PACS numbers: 11.10.Lm, 11.27.+d,  04.60.Cf, 02.30.Ik}
\end{abstract}
\tableofcontents

%%%%%%%%%%%%%%%%%%%%%%%%%%%%%%%%%%%%%%%%%%%%%%%%%%%%%%%%%%%%%%%%%%%%%%%%%%%%%%%
%% Introduction
%%%%%%%%%%%%%%%%%%%%%%%%%%%%%%%%%%%%%%%%%%%%%%%%%%%%%%%%%%%%%%%%%%%%%%%%%%%%%%%

\section{Introduction}
In string theory, it is natural to describe the dynamics of a string propagating in curved background using non-linear sigma model satisfying supplementary conditions. However, finding solutions of equations of motion in these backgrounds is usually very complicated and every solvable case attracts considerable attention.

{Searching} for non-trivial solvable sigma models, one may utilize T-duality, which is known for connecting "dual" models with strikingly different curvature properties \cite{buscher:ssbfe}. Solutions of mutually dual models are related by T-duality transformation. Hence, having solved equations of a model {on a background with isometries}, we may try to use this transformation to find the solution of the dual one. As equations of a sigma model living in flat torsionless background are easily solved, we may ask what are the dual models whose solution can be found via T-duality transformation.

Notions of Abelian and non-Abelian T-duality were introduced in Refs. \cite{buscher:ssbfe} and \cite{delaossa:1992vc} respectively. 
{Global aspects of T-duality, analyzed first in \cite{AAG:global}, 
%lags behind
are still insufficiently understood, resulting in problems %especially 
at the quantum level. That's why we address only local properties of sigma model backgrounds.} The original dualization procedures are based on gauging the symmetries of the background that should be dualized. In this paper, we will use an alternative approach, namely, formulation of non-Abelian T-duality in the framework of \pltd y. It is based on the structure of \dd{} and provides compact formulas for dual backgrounds even for {much more general type of symmetries \cite{klise}. Non-Abelian T-duality is a special case of \pltd y and it is easy to adapt all formulas to this case.

In previous paper \cite{hlapet:cqgrav}, we have classified T-duals of flat metric in four dimensions with respect to four-dimensional subgroups of Poincar\'e group. Such a case for which the dimensions of the isometry group and target manifold are equal is referred to as atomic duality. In this paper, we complete the classification of duals of flat metric in four dimensions by giving its duals with respect to isometry groups whose dimensions are lower than the dimension of the target manifold. In such a case the {so-called} spectator fields appear. Although relevant formulas were given already in {Ref.} \cite{klise}, description of obstacles encountered in this case is, up to our knowledge, missing. We give a thorough discussion of the process of dualization in the presence of spectators {in Sec. \ref{sec2}} and emphasize the differences between duality with spectators and atomic duality.

To find the dual model, we have to introduce the {so-called} adapted coordinates first. These are chosen in such a way as to identify submanifolds invariant under the action of the particular isometry group with the group itself.
%It may be deduced from
The results of paper \cite{hlafilp:uniq} indicate that up to torsionless antisymmetric part of dual backgrounds the non-uniqueness in the choice of adapted coordinates can be interpreted as a change of \coor s in the target manifold of the corresponding dual sigma model. This opens the way for classification of (non-)Abelian T-duals of the flat metric in four dimensions with respect to {one-, two-, and three-dimensional} subgroups of the Poincar\'e group.

As we shall see, similarly to paper \cite{hlapet:cqgrav}, there are
two types of dual backgrounds -- diagonalizable metrics or
plane-parallel (pp-)waves, both with and without torsion. We focus
mainly on the pp-wave backgrounds. {These were repeatedly
investigated in the past in context of string theory since they give
exact string backgrounds {\cite{dhh:spfgw,tsey:revExSol,tseyt94}} or
allow to study the behavior of strings near space-time singularity
\cite{HorSteif,VegaSan}. %For the sake of brevity, we do not analyze
%the resulting dual models in detail. 
An {in-depth} study of
particular cases of pp-wave backgrounds was carried out e.g. in
{Refs. \cite{Sfetseyt94,BLPT,papa}}.  

{The non-Abelian T-duality
 was recently extended to superstring backgrounds including the Ramond fluxes
\cite{1012.1320 sfethomp,1104.5196 LCST} with the supersymmetric  plane-parallel wave as well as supergravity background studied in Refs. \cite{1205.2274 ILCS,Itsios:2013wd,DMRV}. Topological properties of RR-fields were investigated by non-Abelian T-duality in the paper \cite {Gevorgyan:2013xka}% and another supersymmetric  backgrounds in \cite{1212.1043 LCrgS},\cite{1305.7229 BGMNT}
. Applications  of non-Abelian duality in cosmology, more precisely
on FLRW metrics, can be found e.g. in Ref. \cite{0210211 BosMoham}.}}

The structure of the paper is the following. In Sec. 2, we describe in detail the steps
necessary to find dual backgrounds via \pltd y with spectators.
{We introduce adapted coordinates and summarize formulas used later in the paper to construct
dual sigma models.} In Sec. 3, we analyze properties of actions of one-, two-, and
three-dimensional subgroups of Poincar\'e group on the target manifold and define their
invariant submanifolds. In Sec. 4, we list the resulting pp-wave backgrounds in their
standard forms in Brinkmann and Rosen coordinates and give convenient forms of diagonalizable
metrics as well. Concluding remarks are contained in Sec. 5. In order to keep the {paper}
more compact, we do not list dual backgrounds coming directly from the dualization procedure
in the text but summarize them in the Appendix.

\section{T-duality of \sm s with spectators}\label{sec2}
{Non-linear \sm s on a manifold $M$ can be given by a tensor field $\cf$ and action
\begin{equation}\label{sigm1} S_{\cf}[\phi]=\int_\Omega \partial_-
\phi^{\mu}\cf_{\mu\nu}(\phi)
\partial_+ \phi^\nu\,d\xi_+ d\xi_-.
\end{equation}Symmetric and antisymmetric part of the tensor $\cf$ represent metric and torsion potential of the \sm{} background. }

Both Abelian \cite{buscher:ssbfe} and non-Abelian
\cite{delaossa:1992vc} T-dualities can be applied to non-linear
sigma models if tensor field on the target manifold $M$ is invariant
under an action of the Lie group $G$. In other words, if the Lie
group is generated by independent Killing vector fields $K_a,\
a=1,\ldots,{\dim}\,G$ the condition for dualizability of the \sm{}
is
\begin{equation}       \label{killeq}
  (\mathfrak{L}_{K_a}\cf)_{\mu\nu} =0,\ a=1,\ldots,{\dim}\,G
 \end{equation}  where $\mathfrak{L}$ denotes the Lie derivative.

In this paper, we focus on the case where
${\dim}\,G\,<{\dim}\,M$, {and consequently one can not} identify
$G\approx M$ (atomic duality). Nevertheless, we shall assume that
the groups of isometries act freely and transitively on
submanifolds of $M$ invariant under the isometry group.
Therefore, we can identify them with orbits of $G$.

Let us summarize main points of construction of dual models by the \dd{}
method in this case.

\subsection{Invariant submanifolds and adapted coordinates}
{Standard dualization procedures \cite{buscher:ssbfe,delaossa:1992vc,klise} start with assumption} that the initial background is expressed in the \coor s adapted to the action of the symmetry group. However, when we want to dualize a particular background (the flat one in our case), this assumption need not be satisfied. Therefore,
before the dualization procedure is started, we have to construct
invariant submanifolds $\Sigma$ of $M$ in order to introduce adapted \coor s.

The invariant submanifolds are implicitly
given by functions
$\Phi(x^\mu)$ %obtained from the action of Killing vectors on coordinate functions by condition
satisfying linear partial differential
equations\begin{equation}\label{invariantconds}
    K_a\Phi=0,\ \ a=1,\ldots,{\dim}\,G
\end{equation}
where $K_a$ are Killing vector fields of the symmetry group $G$. If the dimension of the manifold $M$ is greater than dimension of the group $G$,
we obtain $S={\dim}\,M\,-{\dim}\,G$ independent solutions
defining the invariant submanifolds $\Sigma(s)$ as
\begin{equation}\label{submanifolds implicitly}
\Phi^\delta(x^\mu)=s^\delta,\ \
 \delta=1,\ldots,S.\end{equation}
Assuming that the action of the isometry group is free, we can identify each
of the invariant submanifolds with the isometry group and Killing
vectors with left-invariant fields of the group. The latter
identification provides us with transformation to adapted
coordinates on $M$ \begin{equation}\label{adapted}
x'^\mu=\{s^\delta,g^a\},\ \
 \delta=1,\ldots,S,\ \ a=1,\ldots,{\dim}\,G
\end{equation} with $s^\delta$ numerating
the invariant submanifolds and $g^a$ parameterizing group elements by
$$g=\e^{g^{1}T_{1}}\e^{g^{2}T_{2}}\ldots\, \e^{g^{\dim G}T_{\dim G}}$$
where $T_a$ form a basis of the Lie algebra of the group $G$.

The left-invariant vector fields $V(x')$ can be extended to $M$ as
$$V^\delta(x')=0,\ \delta=1,\ldots,S$$ and \eqn s determining \tfn s  to the adapted \coor s take the form
\begin{equation}\label{KtoL}
    K_a^\mu(x)=\frac{\partial x^\mu}{\partial x'^\nu}V_a^\nu(x'), \
    a=1,\ldots,{\dim}\,G
\end{equation}where $V_a$ are independent left-invariant fields commuting in
the same way as the corresponding Killing vectors. As the number of equations (\ref{KtoL})
is less than the number of sought functions
$X^\mu(x')=x^\mu$, solution will depend on undetermined functions
$\xi^\mu(s^\delta)$, i.e.
\begin{equation}\label{xmuXmu}
x^\mu=X^\mu_\xi(x')=X^\mu_\xi(s,g).
\end{equation}
Actually, the functions $\xi^\mu$ assign the group unit to points of the submanifolds $\Sigma$. Consequently, the functions $\xi^\mu$ are restricted by condition
(\ref{submanifolds implicitly}) that must be satisfied also in the
unit element of the group so that
\begin{equation}\label{submanifolds in e}
\Phi^\delta(X^\mu_\xi(s,0))=\Phi^\delta(\xi^\mu(s))=s^\delta,\ \
 \delta=1,\ldots,S.\end{equation}
Moreover, the embedded manifold $\xi^\mu(s)$ in $M$ must be
transversal to invariant submanifolds. In other words, its
tangent vectors $\Xi_\delta,\ \delta=1,\ldots,S$ with components
$\Xi_\delta^\mu= \frac{\partial \xi^\mu}{\partial s^\delta}$ and
Killing vector fields $K_a$ in the location of the group unit on the
invariant submanifolds must be independent. It means that for
{all $s$
\begin{equation}\label{transverse}
 \det \left[\,\Xi_\delta^\mu(s),K_a^\mu(\xi(s))\right]\neq 0.
\end{equation}
In fact}, definition of spectators (\ref{submanifolds in e}) provides
\begin{equation}
\frac{\partial \Phi^\alpha}{\partial x^\mu}\left(\xi^\mu(s)\right)\frac{\partial \xi^\mu}{\partial s^\beta}(s)=\delta^\alpha_\beta,
\end{equation}
which means that each of the tangent vectors ${\Xi_\delta}$ exhibits a non-trivial component transversal  to invariant submanifolds defined by $\Phi^\delta$.
Thus, it assures that condition (\ref{transverse}) is satisfied because the action given by the
Killing vectors $K_a$ is transitive on the invariant submanifolds.

To be able to apply formulas for dualization procedure, components of the tensor
field $\cf$ must be expressed in the adapted coordinates
(\ref{adapted}). However, components $F_{\kappa\lambda}$ of the tensor field $\cf$ in
adapted \coor s will depend on functions $\xi^\mu$ since the \tfn{}
matrix $\frac{\partial x^\mu}{\partial x'^\kappa}$
does.\begin{equation}\label{FtoF}
    (F_\xi)_{\kappa\lambda}(s,g)=\frac{\partial X^\mu_\xi(x')}{\partial x'^\kappa}
    \frac{\partial X^\nu_\xi(x')}{\partial
    x'^\lambda}\cf_{\mu\nu}(X_\xi(x')).
\end{equation}
The condition (\ref{killeq}) for  non-Abelian dualizability of the
\sm{} then takes the form
\begin{equation}       \label{kseq0}
  (\mathfrak{L}_{V_{a}}F_\xi)_{\mu\nu} =0.
 \end{equation}
{\subsection{Dualizable tensor fields}}
{As mentioned in the Introduction, (non-)Abelian T-duality can be considered as} a special case of the \pltd y
\cite{klise, klim:proc} formulated in the framework of \dd{}  -- a Lie group whose Lie algebra $\cd$ admits a decomposition
$\cd=\cg\dotplus\wh\cg$ into a pair of subalgebras maximally
isotropic with respect to a symmetric ad-invariant non-degenerate
bilinear form $\langle\, .\,,.\,\rangle_\cd $.

In general, the \pltd y can be applied to models with tensor fields
satisfying \cond{}
\begin{equation}       \label{kseq}
  (\mathfrak{L}_{V_{a}}F)_{\mu\nu} =
   F_{\mu\kappa}V^{\kappa}_{b}\wh{f}^{bc}_{a} %krat
   V^{\lambda}_{c}F_{\lambda\nu}
 \end{equation}
where $V_b^\kappa$ are components of the left-invariant vector
fields $V_b$ generating a Lie group $G$ and $\wh{f}^{bc}_{a}$ are
 structure coefficients of {the Lie algebra of a "dual" Lie group $\wh G$ of the same dimension as $G$.} Self-consistency of the
 condition (\ref{kseq}) restricts the structure coefficients in such way
 that $G$ and $\wh G$ can be interpreted
 as subgroups defining the \dd{} $D\equiv(G|\wh G)$.

{Components of the tensor field $\cf$ satisfying the \cond{}
(\ref{kseq}) %with $\wh{f}^{bc}_{a}=0$
can be written as
\begin{equation}\label{met}
        (F_\xi)_{\mu\nu}(s,g)=e_{\mu}^{j}(g)\left[\left(E_\xi(s)\-1+\Pi(g)\right)\-1\right]_{jk}e_{\nu}^{k}(g)
    \end{equation} {with matrix $E_\xi(s)=F_\xi(s,0)$ independent of the group \coor s  and
$$
e_{\mu}^{j}(g)=\left(\begin{array}{cc} \unit_S & 0 \\ 0
&e_{\mu}^{a}(g)
     \end{array}\right),\ \ \Pi(g)=\left(\begin{array}{cc} \mathbf O_S & 0 \\ 0
&b(g)\cdot a(g)\-1
     \end{array}\right) $$
where $e_{\mu}^{a}(g)$ are components} of right-invariant
forms
$(dg)g^{-1}$  expressed %)_{\mu}^{a}$
in group coordinates,
$a(g),\ b(g)$ are matrices obtained from the adjoint representation of the  group $G$ on the algebra $\cd$ and
$$ e_{\mu}^{a}(0)=\unit_G,\ \ a^{j}_{k}(0)=\unit_G,\ \ b_{jk}(0)=\mathbf O_G.$$
Matrices $\unit_G,\unit_S $ and $\mathbf O_G,\mathbf O_S$ are unit
and zero matrices of ${\dim}\,G$ and $({\dim}\,M-{\dim}\,G)$. In
case of (non-)Abelian T-duality $\wh G$ is Abelian,
$\wh{f}^{bc}_{a}=0$,} $a(g)=\unit_G,\ \ b(g)= \mathbf O_G$, resulting in
\begin{equation}\label{met0}
        (F_\xi)_{\mu\nu}(s,g)=e_{\mu}^{j}(g)E_\xi(s)_{jk}e_{\nu}^{k}(g).
    \end{equation}
%we obtain the \cond{} (\ref{kseq0}).

\emph{ Fortunately,  we
do not have to solve the \eqn s (\ref{KtoL}) for $X^\mu_\xi$ to obtain
$F_\xi$ and compute matrix $E_\xi(s)$ from (\ref{met}).} Instead, we can use the fact that
\begin{equation}\label{Exi}
(F_\xi)_{\kappa\lambda}(s,0)=\frac{\partial X^\mu_\xi}{\partial
x'^\kappa}(s,0)
    \frac{\partial  X^\nu_\xi}{\partial
    x'^\lambda}(s,0)\cf_{\mu\nu}(X_\xi(s,0))
\end{equation}
with $X^\mu_\xi(s,0)=\xi^\mu(s),$ and
\begin{equation}\label{derces in e}
    \frac{\partial X^\mu_\xi}{\partial
s^\delta}(s,0)=\frac{\partial \xi^\mu}{\partial s^\delta}(s).
\end{equation}
Moreover, due to (\ref{KtoL})\begin{equation}\label{derceg in e}
    \frac{\partial X^\mu_\xi}{\partial
g^a}(s,0)=K_b^\mu(\xi(s))(V\-1)_a^b(0)=K_a^\mu(\xi(s)).
\end{equation}
Summarizing (\ref{Exi})-(\ref{derceg in e}), we can get $ E_\xi(s)$ in the
block form as
\begin{equation}        \label{Exi blok}
 E_\xi(s)=\left(\begin{array}{cc}
    E_{\alpha\beta}(s) & E_{\alpha d}(s) \\
     E_{c\beta}(s)      &E_{cd}(s)    \end{array}\right),\
     \ \alpha,\beta=1,\ldots,S,\ \ c,d=1,\ldots,\text{dim}\,G\ \
\end{equation} where
\begin{eqnarray}\nonumber
   E_{\alpha\beta}(s) &=& \frac{\partial \xi^\mu}{\partial s^\alpha}(s)
   \, \cf_{\mu\nu}(\xi(s))\,\frac{\partial \xi^\nu}{\partial s^\beta}(s),\\ \label{Exicomps}
  E_{\alpha d}(s) &=&  \frac{\partial \xi^\mu}{\partial s^\alpha}(s)
   \, \cf_{\mu\nu}(\xi(s))\,K_d^\nu(\xi(s)), \\\nonumber
  E_{c \beta}(s) &=&
   K_c^\mu(\xi(s))\, \cf_{\mu\nu}(\xi(s))\,\frac{\partial \xi^\nu}{\partial s^\beta}(s), \\\nonumber
 E_{cd}(s)&=& K_c^\mu(\xi(s))
    \,\cf_{\mu\nu}(\xi(s))\,K_d^\nu(\xi(s))
\end{eqnarray}
and $\cf_{\mu\nu}$ are components of $\cf $ in any
\coor s.

\subsection{Dual tensor field}
{Components of the dual tensor $\wh\cf$ obtained by the \pltd y with spectators \cite{klim:proc,hlapevoj} 
 are
\begin{equation}\label{Fghat1} (\wh F_\xi)_{\mu\nu} (s,\ghat)=\wh e_{\mu}^{j}(\ghat)\left[\left(\wh E_\xi(s)\-1+\wh \Pi(\ghat)\right)\-1\right]_{jk}\wh e_{\nu}^{k}(\ghat)
\end{equation}
where  $\wh e_{\mu}^{j}(\ghat)$ contain components of right-invariant
forms on $\wh G$, $\wh a(\ghat),\ \wh b(\ghat)$ are matrices obtained from the adjoint representation of the dual group $\wh G$ on the algebra $\cd$. Matrix  $\widehat E_\xi(s)$ is given by  \begin{equation}\label{Extil}
  \wh E_\xi(s)=\big(A+E_\xi(s)\cdot{B}
    \big)^{-1}\cdot\big(B+E_\xi(s)\cdot A\big)
\end{equation}
where\begin{equation}\label{matsAB}
{A}=\left(\begin{array}{cc} \unit_S & 0 \\ 0 & \mathbf O_G
     \end{array}\right), \quad
{B}=\left(\begin{array}{cc}  \mathbf O_S & 0 \\ 0 & \unit_G
     \end{array}\right). \end{equation}
{Using formulas \eqref{Fghat1}-\eqref{matsAB} and \eqref{Exi blok}, we get $ \wh F_\xi(s,\ghat)$ in the
block form as
\begin{equation}        \label{Fxi blok}
 \wh F_\xi(s,\ghat)=\left(\begin{array}{cc}
    \wh F_{\alpha\beta}(s,\ghat) & \wh F_{\alpha d}(s,\ghat)\wh e_{\nu}^{d}(\ghat) \\
    \wh e_{\mu}^{c}(\ghat) \wh F_{c\beta}(s,\ghat)      &\wh e_{\mu}^{c}(\ghat)\wh F_{cd}(s,\ghat)\wh e_{\nu}^{d}(\ghat)    \end{array}\right)\end{equation} where
\begin{eqnarray}\nonumber
   \wh F_{\alpha\beta}(s,\ghat) &=& E_{\alpha\beta}(s)-E_{\alpha d}(s)(M(s,\ghat)\-1)_{dc}E_{c\beta}(s),\\ \label{Fxicomps}
  \wh F_{\alpha d}(s,\ghat) &=& -E_{\alpha c}(s)(M(s,\ghat)\-1)_{cd}, \\\nonumber
  \wh F_{c \beta}(s,\ghat) &=&
  (M(s,\ghat)\-1)_{cd}E_{d\beta}(s)), \\ \nonumber
 \wh F_{cd}(s,\ghat)&=&(M(s,\ghat)\-1)_{cd},\\
  M_{cd}(s,\ghat)&=&E_{cd}(s)+\wh \Pi_{cd}(\ghat)\label{Mmat}.
\end{eqnarray}
{Dual metric and torsion potential then can be obtained as symmetric and antisymmetric parts of expressions \eqref{Fxicomps}.}}

{In case of (non-)Abelian T-duality, %\tfn{} of (\ref{met}),
the dual group is Abelian, $\wh e_{\mu}^{c}(\ghat)=\delta_\mu^c$ and
\begin{equation} \label{Pihat2}
\wh \Pi(\hat g)=\left(\begin{array}{cc}  \mathbf O_S & 0 \\ 0 & -{f_{cd}}^{b}\ghat_b%\wh b(\hat g)
  \end{array}\right)
\end{equation}
where ${f_{cd}}^{b}$ are structure coefficients of the Lie algebra
of the group  $G$ and $\ghat_b$ are coordinates of the Abelian group
$\wh G$. Equations \eqref{Fxi blok}-\eqref{Mmat} then become the
Buscher formulas for non-Abelian T-duality (cf.
\cite{Itsios:2013wd,DMRV,Gevorgyan:2013xka}).}
The dual tensor \eqref{Fghat1} in this case can be written as}
\begin{equation}\label{Fghat2} \wh F_\xi (s,\ghat)=(\unit+\wh
E_\xi(s)\cdot\wh \Pi(\ghat))^{-1}\cdot \wh E_\xi(s).
\end{equation}
As the first factor in (\ref {Extil}) of $\wh E_\xi(s)$
must have non-vanishing determinant, we get from (\ref{Exicomps}),
(\ref{Extil}), and (\ref{matsAB}) condition\footnote{{Note that the
formula \eqref{Fghat2} {used  here}
%in the presence of spectators,}
differs from \eqref{Fghat1}.
% $\wh F_\xi (s,\ghat)=(\wh E_\xi(s)^{-1}+\wh b(\ghat))^{-1}$
%used in Ref.\cite{hlapevoj}.
{They are equivalent for
invertible $E_\xi(s)$}.} To be able to compute $\wh E_\xi(s)^{-1}$,
we should also require $\det (E_{\alpha\beta}(s))\neq 0$ to ensure
that the second term in \eqref{Extil} is invertible, which is not
necessary now.}
\begin{equation}\label{condsAB}
\det\, \left(A+E_\xi(s)\cdot{B} \right)=\det\, \left(E_{cd}(s)\right)\neq 0
\end{equation}
{which further restricts} the \fn s $\xi^\mu$.

Using formulas \eqref{Extil},(\ref{Pihat2}), and (\ref{Fghat2}) we gain dual tensors
whose components may depend on functions $\xi^\mu$.
% or $X_\xi$ %:x^\mu\mapsto X^\mu_\xi (s^\delta,g^b)\in \Sigma(s)$mapping for fixed $s$ from the group $G$ to the invariant manifold $\Sigma(s)$.
As we have suggested in Ref. \cite{hlafilp:uniq}, dual
backgrounds with different \fn s $\xi^\mu$ can be obtained by \coor{}
\tfn s in the target manifolds $M$ and $\wh M$. Up to \cond s
(\ref{submanifolds in e}) and (\ref{condsAB}), we can choose these functions arbitrarily.

\section {Poincar\'e subalgebras}\label{PSA}
It is clear from (\ref{Exicomps})-(\ref{Pihat2}) that the basic
data for \dualn{} are generators  of the Lie group $G$. In our case, they are the Killing vectors of the flat metric
%spacetime coordinates
$\eta=\text{diag}(-1,1,1,1)$, which is to be dualized.
The generators are
\begin{equation}\label{Poincar\'e killings}
    P_0=\partial_t,\ P_j=\partial_j,\ L_j=-\varepsilon_{jkl}x^k\partial_l,\
    M_j=-x^j\partial_t-t\partial_j
\end{equation} and form ten-dimensional Poincar\'e Lie algebra.  Its one-, two- and three-dimensional
subalgebras classified up to conjugacy under the proper
orthochronous Poincar\'e group are given in Ref. \cite{PWZ}. {For every subalgebra in the classification, we must find
open subsets of $\real^4$ where the action is  transitive and free and explicit forms of \fn s
$\Phi^\delta$ defining invariant submanifolds. Bellow in {Tabs.}
\ref{table1}-\ref{table3}, we summarize the results of these
calculations.}
\begin{table}
\begin{center}
\scriptsize {\renewcommand{\arraystretch}{1.4}
\begin{tabular}{|c || c | c | c |}
\hline 
 & Generators & Free and transitive action for &  $\Phi^1(x^{\mu}), \Phi^2(x^{\mu}), \Phi^3(x^{\mu})$   \\
\hline \hline $S_{1,1}$ & $\begin{array}{c}
\cos \gamma L_3 + \sin\gamma M_3, \\
0 < \gamma < \pi, \gamma \neq \frac{\pi}{2}
\end{array}$ & $\begin{array}{c} t\neq 0 \lor x \neq 0 \ \lor \\ \lor\ y\neq 0 \lor z \neq 0 \end{array}$ & $\begin{array}{c} -t^2+z^2, \\ x \cos (f_\gamma(t+z))+y \sin(f_\gamma(t+z)),\\ y \cos(f_\gamma(t+z))-x \sin(f_\gamma(t+z)), \\ \text{where}\, f_\gamma(x)=\cot(\gamma)\ln|x| \end{array}$  \\
\hline
$S_{1,2}$ &  $L_3$ & $x\neq 0 \lor y \neq 0$ & $t,\ z,\ x^2+y^2$  \\
\hline
$S_{1,3}$ & $M_3$ & $t\neq 0 \lor z \neq 0$ & $x,\ y,\ -t^2+z^2$  \\
\hline
$S_{1,4}$ & $L_2 + M_1$ & $x\neq 0 \lor t+z \neq 0$ & $t+z,\ \frac{1}{2}\left(-2 t^2+x^2-2 t z\right),\ y$  \\
\hline
$S_{1,5}$ & $P_0 - P_3$ & $\mathbb{R}^4$ & $t+z,\ x,\ y$  \\
\hline
$S_{1,6}$ & $P_3$ & $\mathbb{R}^4$ & $t,\ x,\ y$  \\
\hline
$S_{1,7}$ & $P_0$ & $\mathbb{R}^4$ & $x,\ y,\ z$  \\
\hline
$S_{1,8}$ & $\begin{array}{c} L_3 + \epsilon (P_0 + P_3),\\ \epsilon = \pm 1  \end{array}$ & $\mathbb{R}^4$ & $\begin{array}{c} -t+z,\\ x \cos\left(\frac{t}{\epsilon }\right)-y \sin\left(\frac{t}{\epsilon }\right),\\ y \cos\left(\frac{t}{\epsilon }\right)+x \sin\left(\frac{t}{\epsilon }\right) \end{array}$  \\
\hline
$S_{1,9}$ & $\begin{array}{c} L_3 + \alpha P_0,\\ \alpha > 0 \end{array}$ & $\mathbb{R}^4$ & $\begin{array}{c} z,\\  y \cos\left(\frac{t}{\alpha }\right)+x \sin\left(\frac{t}{\alpha }\right),\\ x \cos\left(\frac{t}{\alpha }\right)-y \sin\left(\frac{t}{\alpha }\right) \end{array}$  \\
\hline
$S_{1,10}$ & $\begin{array}{c} L_3 + \alpha P_3,\\ \alpha \neq 0 \end{array} $ & $\mathbb{R}^4$ & $\begin{array}{c} t,\ x^2+y^2,\\ z-\alpha\, \text{arctan}\left(\frac{x}{y}\right) \end{array}$  \\
\hline
$S_{1,11}$ & $\begin{array}{c} M_3 + \alpha P_1,\\ \alpha > 0 \end{array}$ & $\mathbb{R}^4$ & $y,\ -t^2+z^2,\ x-\alpha  \log |t-z|$  \\
\hline
$S_{1,12}$ & $L_2 + M_1 + P_0 + P_3$ & $\mathbb{R}^4$ & $\begin{array}{c} y,\  4 x+(t+z)^2, \\ 6(z-t)-(t+z)(6 x+(t+z)^2)\end{array}$ \\
\hline
$S_{1,13}$ & $\begin{array}{c} L_2 + M_1 + \epsilon P_2, \\ \epsilon = \pm 1\end{array}$ & $\mathbb{R}^4$ & $\begin{array}{c} t+z,\ \frac{1}{2} \left(-2 t^2+x^2-2 t z\right),\\  y+\frac{x \epsilon }{t+z} \end{array}$ \\
\hline
\end{tabular}
} \normalsize \normalsize \caption {{Open subsets of transitive and
free action and invariant functions for one-dimensional subgroups of
the Poincar\'e group}\label{table1}}
\end{center}
\end{table}

\begin{table}
\begin{center}
\scriptsize {\renewcommand{\arraystretch}{1.4}
\begin{tabular}{|c || c | c | c |}
\hline
%Subgroup
 & Generators & Free and transitive action for & $\Phi^1(x^{\mu}), \Phi^2(x^{\mu})$ \\
\hline \hline
$S_{2,1}$ &  $L_3, M_3$ & $\begin{array}{c} y z \neq 0 \lor x z \neq 0\ \lor \\
\lor\ t y \neq 0 \lor t x \neq 0\end{array}$ & $t^2 - z^2,\ x^2 + y^2$ \\
\hline
$S_{2,2}$ & $L_2 + M_1, L_1-M_2$ & $t+z\neq 0$  & $t + z,\ -2 t^2 + x^2 + y^2 - 2 t z$ \\
\hline
$S_{2,3}$ & $L_3, P_0-P_3$ & $x\neq 0 \lor y\neq 0$ & $t+z,\ x^2 + y^2$\\
\hline
$S_{2,4}$ & $L_3, P_3$ & $x\neq 0 \lor y\neq 0$ & $t,\ x^2 + y^2$ \\
\hline
$S_{2,5}$ & $L_3, P_0$& $x\neq 0 \lor y\neq 0$ & $z,\ x^2 + y^2$ \\
\hline
$S_{2,6}$ & $M_3, P_1$ & $t\neq 0 \lor z\neq 0$ & $t^2 - z^2,\ y$ \\
\hline
$S_{2,7}$ & $L_2 + M_1, P_0 - P_3$ & $t+z\neq 0$ & $t+z,\ y$ \\
\hline
$S_{2,8}$ & $L_2 + M_1, P_2$ & $t+z\neq 0 \lor x\neq 0$ & $t + z,\ -2 t^2 + x^2 - 2 t z$ \\
\hline
$S_{2,9}$ & $P_0-P_3, P_1$ & $\mathbb{R}^4$ & $t+z, \ y$ \\
\hline
$S_{2,10}$ & $P_0, P_3$& $\mathbb{R}^4$ & $x,\ y$ \\
\hline
$S_{2,11}$ & $P_1, P_2$ & $\mathbb{R}^4$ & $t,\ z$ \\
\hline $S_{2,12}$ & $\begin{array}{c} L_2 + M_1, L_1-M_2+P_2
\end{array}$ & $\begin{array}{c}
y (t+z)\neq 0\  \lor \\ \lor\ x (1+t+z)\neq 0\ \lor \\
\lor\ (t+z) (1+t+z)\neq 0
\end{array}$ & $t+z,\ \frac{x^2-(t+z) \left(-x^2-y^2+2 t (1+t+z)\right)}{2 (t+z)}$ \\
\hline
$S_{2,13}$ & $\begin{array}{c} L_3+\epsilon(P_0+P_3), P_0-P_3,\\ \epsilon = \pm 1 \end{array}$ & $\mathbb{R}^4$ & $\begin{array}{c} y \cos\left(\frac{t+z}{2 \epsilon }\right)+x \sin\left(\frac{t+z}{2 \epsilon }\right), \\ x \cos\left(\frac{t+z}{2 \epsilon }\right)-y \sin\left(\frac{t+z}{2 \epsilon }\right) \end{array}$ \\
\hline
$S_{2,14}$ & $\begin{array}{c} L_3+\alpha P_0, P_3,\\ \alpha>0 \end{array}$ & $\mathbb{R}^4$ & $\begin{array}{c} y \cos\left(\frac{t}{\alpha }\right)+x \sin\left(\frac{t}{\alpha }\right),\\ x \cos\left(\frac{t}{\alpha }\right)-y \sin\left(\frac{t}{\alpha }\right) \end{array}$ \\
\hline
$S_{2,15}$ & $\begin{array}{c}L_3+\alpha P_3, P_0,\\ \alpha\neq 0 \end{array}$ & $\mathbb{R}^4$ & $x^2+y^2,\ -z+\alpha\, \text{arctan}\left(\frac{x}{y}\right)$ \\
\hline
$S_{2,16}$ & $\begin{array}{c} M_3+\alpha P_2, P_1,\\ \alpha>0 \end{array}$ & $\mathbb{R}^4$ & $-t^2+z^2,\ y-\alpha  \ln|t-z|$ \\
\hline
$S_{2,17}$ & $\begin{array}{c} L_2+M_1+P_0+P_3, P_0-P_3 \end{array}$ & $\mathbb{R}^4$ & $x+\frac{1}{4} (t+z)^2,\ y$ \\
\hline
$S_{2,18}$ & $\begin{array}{c} L_2+M_1+\epsilon P_2, P_0-P_3,\\ \epsilon = \pm 1 \end{array}$ & $\mathbb{R}^4$ & $t+z,\ y+\frac{x \epsilon }{t+z}$\\
\hline
$S_{2,19}$ & $\begin{array}{c} L_2+M_1+P_0+P_3, P_2\end{array}$ & $\mathbb{R}^4$ & $\begin{array}{c} 4 x+(t+z)^2, \\ 6 z-6 t-t^3-6 t x-3 t^2 z-6 x z-3 t z^2-z^3 \end{array}$ \\
\hline
$S_{2,20}$ & $M_3, L_2+M_1$ & $\begin{array}{c} z (t+z)\neq 0\ \lor \\ \lor\ x (t+z)\neq 0\ \lor \\
\lor\ t (t+z)\neq 0\end{array}$ & $t^2-x^2-z^2,\ y$ \\
\hline $S_{2,21}$ & $\begin{array}{c} \cos \gamma L_3 + \sin\gamma
M_3, P_0-P_3,\\ 0 < \gamma < \pi, \gamma \neq \frac{\pi}{2}
\end{array}$ & $\begin{array}{c} t+z\neq 0\ \lor \\ \lor\ x\neq 0
\lor y\neq 0 \end{array}$ &  $\begin{array}{c}
 -y \cos(f_\gamma(t+z))+x \sin(f_\gamma(t+z)), \\
 x \cos(f_\gamma(t+z))+y \sin(f_\gamma(t+z)), \\
 \text{where } f_\gamma(x)=\cot(\gamma)\ln|x| \end{array}$ \\
\hline
$S_{2,22}$ & $M_3, P_0-P_3$ & $t+z\neq 0$ & $x,\ y$ \\
\hline $S_{2,23}$ & $\begin{array}{c} M_3+\alpha P_2, L_2+M_1,\\
\alpha>0 \end{array}$ & $t+z\neq 0 \lor x\neq 0$ & $\begin{array}{c}
-t^2+x^2+z^2,\ y+\alpha  \ln|t+z|
\end{array}$ \\
\hline
$S_{2,24}$ & $\begin{array}{c} M_3+\alpha P_2, P_0-P_3,\\ \alpha>0 \end{array}$ & $\mathbb{R}^4$ & $x,\ e^{\frac{y}{\alpha }} (t+z)$ \\
\hline
\end{tabular}
\normalsize \caption { {Open subsets of transitive and
free action and invariant functions for two-dimensional subgroups of
the Poincar\'e group}\label{table2}} }
\end{center}
\end{table}

\begin{table}
\begin{center}
\scriptsize{\renewcommand{\arraystretch}{1.4}
\begin{tabular}{|c||c|c|c|}
\hline
%Sub-group
& Generators & Free and transitive action for &  $\Phi^1(x^\mu)$ \\
\hline \hline
$S_{3,1}$ &  $L_2+M_1, L_1-M_2, P_0-P_3$ & $t+z\neq0$ & $t+z$ \\
\hline
$S_{3,2}$ & $L_3, P_0, P_3$ & $x\neq 0 \lor y\neq 0$ & $x^2+y^2$ \\
\hline
$S_{3,3}$ & $M_3, P_1, P_2$ & $t\neq 0 \lor z\neq 0$ & $t^2-z^2$ \\
\hline
$S_{3,4}$ & $L_2+M_1, P_0-P_3, P_2$ & $t+z\neq0$ & $t+z$ \\
\hline
$S_{3,5}$ & $P_0-P_3, P_1, P_2$& $\mathbb{R}^4$ & $t+z$ \\
\hline
$S_{3,6}$ & $P_1, P_2, P_3$ & $\mathbb{R}^4$ & $t$ \\
\hline
$S_{3,7}$ & $P_0, P_1, P_2$ & $\mathbb{R}^4$ & $z$ \\
\hline
$S_{3,8}$ & $L_2+M_1, L_1-M_2+P_2, P_0-P_3$ & $(t+z)(1+t+z)\neq0$ & $t+z$ \\
\hline
$S_{3,9}$ & $L_2+M_1-\frac{1}{2}(P_0+P_3), P_0-P_3, P_2$ & $\mathbb{R}^4$ & $(t+z)^2-2x$ \\
\hline
$S_{3,10}$ & $M_3, L_2+M_1, P_2$ & $t+z\neq0$ & $t^2-x^2-z^2$ \\
\hline
$S_{3,11}$ & $M_3, P_0-P_3, L_3$ & $(t+z)(x^2+y^2)\neq0$ & $x^2+y^2$ \\
\hline
$S_{3,12}$ & $M_3, P_0-P_3, P_2$ & $t+z\neq0$ & $x$ \\
\hline
{$S_{3,13}$} & $M_3 + \alpha P_2, P_0-P_3, P_1,\  \alpha > 0$ &{$\mathbb{R}^4$} & {$(t+z)\text{e}^\frac{y}{\alpha}$} \\
\hline
$S_{3,14}$ & $L_2+M_1, P_1, P_0-P_3$ & $\emptyset$ & \\
\hline
{$S_{3,15}$} & $L_2+M_1, P_2+\beta P_1, P_0-P_3,\ \beta\neq 0$ & {$t+z\neq0$} &{$t+z$}  \\
\hline
\multirow{2}{*}{$S_{3,16}$} & $L_2+M_1-\epsilon P_2, L_1-M_2+\beta P_2-\epsilon P_1, P_0-P_3,$ & \multirow{2}{*}{$\epsilon^2 + (t+z)(\beta+t+z)\neq0$} & \multirow{2}{*}{$t+z$} \\
 & $\beta > 0, \epsilon = \pm 1$ & & \\
\hline
$S_{3,17}$ & $L_2+M_1-\epsilon P_2, L_1-M_2-\epsilon P_1, P_0-P_3$, $\epsilon = \pm 1$ &$\mathbb{R}^4$ &$t+z$  \\
\hline
$S_{3,18}$ & $L_2+M_1-\frac{1}{2}(P_0+P_3), P_1, P_0-P_3$ & $\mathbb{R}^4$ & $y$ \\
\hline
{$S_{3,19}$} & $L_2+M_1-\epsilon P_2,P_1,P_0-P_3,\ \epsilon = \pm 1$ & {$\mathbb{R}^4$} & {$t+z$} \\
\hline
$S_{3,20}$ & $L_2+M_1-\frac{1}{2}(P_0+P_3), P_2-\beta P_1, P_0-P_3$, $\beta \neq 0$ & $\mathbb{R}^4$ & $(t+z)^2-2(x+\beta y)$ \\
\hline
\multirow{2}{*}{$S_{3,21}$} & $L_2+M_1-\epsilon P_2, P_2 - \beta P_1, P_0-P_3,$ & \multirow{2}{*}{$t+z+\beta\epsilon\neq0$} & \multirow{2}{*}{$t+z$} \\
& $\beta \neq 0, \epsilon = \pm 1$ & & \\
\hline
{$S_{3,22}$} & $M_3 + \alpha P_1, L_2+M_1,P_0-P_3,\ \alpha > 0$ & {$t+z\neq0$} & {$y$} \\
\hline
\multirow{2}{*}{$S_{3,23}$} & $M_3 - \alpha P_2+\beta P_1, L_2+M_1, P_0-P_3,$ & \multirow{2}{*}{$t+z\neq0$} & \multirow{2}{*}{$(t+z)\text{e}^{-\frac{y}{\alpha}}$} \\
 & $\alpha > 0,\beta \neq 0$ & & \\
\hline
$S_{3,24}$ & $M_3, L_2+M_1,L_1-M_2$ & $t+z\neq0$ & $t^2-x^2-y^2-z^2$ \\
\hline
$S_{3,25}$ & $M_3, L_2+M_1, P_0-P_3$ & $t+z\neq0$ & $y$ \\
\hline
{$S_{3,26}$} & $M_3+\alpha P_2, L_2+M_1, P_0-P_3,\ \alpha>0$ & {$t+z\neq0$} & {$(t+z)\text{e}^\frac{y}{\alpha}$} \\
\hline
$S_{3,27}$ & $\cos \gamma L_3 + \sin\gamma M_3, P_0, P_3$, $0 < \gamma < \pi, \gamma \neq \frac{\pi}{2}$ & $x\neq 0 \lor y\neq 0$ & $x^2+y^2$ \\
\hline
$S_{3,28}$ & $M_3, P_0, P_3$ & $\emptyset$  & \\
\hline
{$S_{3,29}$} & $M_3+\alpha P_2, P_0, P_3,\ \alpha>0$ & {$\mathbb{R}^4$} & {$x$} \\
\hline
$S_{3,30}$ &  $L_3, L_2+M_1, L_1-M_2$ & $\emptyset$ & \\
\hline
{$S_{3,31}$} & $\cos \gamma L_3 + \sin\gamma M_3, P_1, P_2,\ 0 < \gamma < \pi, \gamma \neq \frac{\pi}{2}$ & {$t\neq 0 \lor z\neq 0$} & {$t^2-z^2$}  \\
\hline
$S_{3,32}$ &  $L_3, P_1, P_2$ & $\emptyset$ & \\
\hline
{$S_{3,33}$} &  $L_3+\epsilon(P_0-P_3), L_2+M_1, L_1-M_2,\ \epsilon = \pm 1$ & {$t+z\neq0$} & {$t+z$} \\
\hline
{$S_{3,34}$} &  $L_3-\epsilon(P_0+P_3), P_1, P_2,\ \epsilon = \pm 1$ & {$\mathbb{R}^4$} & {$t-z$} \\
\hline
{$S_{3,35}$} &  $L_3+\alpha P_0, P_1, P_2,\ \alpha >0$ & {$\mathbb{R}^4$} & {$z$} \\
\hline
{$S_{3,36}$} &  $L_3+\alpha P_3, P_1, P_2,\ \alpha \neq 0$ & {$\mathbb{R}^4$} & {$t$} \\
\hline
\multirow{2}{*}{$S_{3,37}$} & $\cos \gamma L_3 + \sin\gamma M_3, L_2+M_1, L_1-M_2,$ & \multirow{2}{*}{$t+z\neq0$} & \multirow{2}{*}{$t^2-x^2-y^2-z^2$}  \\
 & $0 < \gamma < \pi, \gamma \neq \frac{\pi}{2}$ &  &  \\
\hline
$S_{3,38}$ & $L_3, M_1, M_2$ & $\emptyset$ &  \\
 \hline
$S_{3,39}$ & $L_1, L_2, L_3$ & $\emptyset$  &  \\
\hline
\end{tabular}
\normalsize \caption{Open subsets of transitive and
free action and invariant functions for three-dimensional subgroups of
the Poincar\'e group} \label{table3}}
\end{center}
\end{table}

\newpage
\section{Results of \dualn}
Using formulas (\ref{Pihat2})-(\ref{Fghat2}), we get dual tensors
{with components depending on functions}
$\xi(s)=(\xi^0(s^\delta),\xi^1(s^\delta),\xi^2(s^\delta),\xi^3(s^\delta)),\
\delta = 1,\ldots,S$. Fortunately,  paper
\cite{hlafilp:uniq} indicates that up to torsionless antisymmetric part, dual backgrounds with different \fn s $\xi$
can be obtained by \coor{} \tfn s in the target manifolds $M$ and
$\wh M$. {It means that up to conditions \eqref{submanifolds in e} and \eqref{condsAB}, we can choose functions
$\xi^\mu(s)$ arbitrarily. We take their particular forms to get simple
expressions of dual backgrounds displayed in the Appendix together with
the corresponding choices of $\xi$. }

As one can see by inspection of results in the Appendix, several
backgrounds obtained by T-duality are again flat metrics. {Furthermore, there} are dual tensors that are of diagonal
form or at least their symmetric part,
i.e. metric, is diagonal, but their
scalar curvature does not vanish. They can be torsionless or {with} torsion. {Besides that, several}
symmetric parts of the dual tensors with non-vanishing scalar curvature can be diagonalized.
Finally, curved dual backgrounds with vanishing
scalar curvature deserve special attention. They turn out to be plane-parallel waves.

\subsection{Plane-parallel waves}

Curved dual backgrounds with vanishing scalar curvature follow from
algebras\begin{equation}\label{S1,x}S_{1,4},S_{1,13},\end{equation}
\begin{equation}\label{s^2,x}S_{2,2},S_{2,8},S_{2,12},S_{2,13},S_{2,21},\end{equation}
\begin{equation}\label{S3,x}S_{3,22},S_{3,23},S_{3,25},S_{3,26}.\end{equation}

These backgrounds turn out to be plane-parallel (pp-)waves
repeatedly investigated in string theory (see e.g. references in {Ref.} \cite{tsey:revExSol} and more recent \cite{Sfetseyt94,BLPT,papa}. The metrics are usually expressed in {the so-called}
Brinkmann {coordinates} as
\begin{equation}\label{BrinkMetrics}
   ds^{2}= 2dudv+\sum_{i,j=3}^4 K_{ij}(u)z_iz_j\,du^2+d{z_3}^2+d{z_4}^2,
\end{equation}
or Rosen \coor s as
\begin{equation}\label{RosenMetrics}
   ds^{2}= 2dudv+\sum_{i,j=1}^2 C_{ij}(u)dx_idx_j.
\end{equation}
{These metrics admit covariantly constant null Killing vector
$\partial _v$ and exhibit particularly simple curvature properties since
the Ricci tensor has only one non-zero component $R_{uu}$ and the
scalar curvature vanishes.
Due to these properties, pp-waves supported by a torsion and dilaton provide exact solutions \cite{dhh:spfgw,tsey:revExSol,tseyt94} of vanishing beta function \eqn s
\begin{eqnarray}
\label{bt1} 0 & = &
R_{\mu\nu}-\bigtriangledown_\mu\bigtriangledown_\nu\phi-
\frac{1}{4}H_{\mu\kappa\lambda}{H_\nu}^{\kappa\lambda},
\\
 \label{bt2} 0 & = & \bigtriangledown^\mu\phi H_{\mu\kappa\lambda}+\bigtriangledown^\mu H_{\mu\kappa\lambda}\,,
\\
\label{bt3} 0 & = & R-2\bigtriangledown_\mu\bigtriangledown^\mu\phi-
\bigtriangledown_\mu\phi\bigtriangledown^\mu\phi-
\frac{1}{12}H_{\mu\kappa\lambda}H^{\mu\kappa\lambda}
\end{eqnarray}
where $H=dB$ is the torsion and $\phi$ is so called dilaton scalar
field.}

{All pp-waves obtained below have the form of the
Penrose--{G{\"u}ven} limit \cite{Penrose,Gueven} with metric
\eqref{BrinkMetrics} and torsion
\begin{equation}\label{Gueven_torsion}
    H=H(u)\,du\wedge dz_3\wedge dz_4.
\end{equation}
The one-loop conformal invariance conditions \eqref{bt1}-\eqref{bt3}
in this case simplify substantially to solvable differential equation
for the dilaton $\phi=\phi(u)$\begin{equation}\label{dilaton_eqn}
   \phi''(u)+K_{33}(u)+K_{44}(u)+\frac{1}{2} H^2(u)=0.%H_{ij}(u)=0.
\end{equation}
}

{Special pp-wave backgrounds obtained in {Refs.
\cite{tsey:revExSol,tseyt94,Sfetseyt94}} from gauged WZW models
are} given in the Rosen coordinates as
\begin{equation}\label{sfets} ds^2= dudv + \frac{g_1 (u')}{ g_1 (u')g_2 (u)  + q^2 }\ dx_1^2 +\frac {g_2 (u)}{ g_1 (u')g_2(u)  + q^2} \ dx_2^2,
\end{equation}
$$B_{12}= \frac{q }{  g_1 (u') g_2 (u) + q^2}$$ where $q= const,\  u'=
au + d \ $ ($a,d=const$) and the functions $g_i{(u)}$ can take {any of the following forms}
\begin{equation}\label{g1g2}
\left\{ 1 ,\, u^2 ,\, \tanh^2 u ,\, \tan^2 u ,\,
 u^{-2} ,\, \coth^2 u ,\, \cot^2 u \right\}.
\end{equation}
Several of these pp-waves {have been} discovered in {Ref.}
\cite{hlapet:cqgrav} to be non-Abelian T-duals of the flat metric,
namely, backgrounds where the functions $g_1, g_2$ are either
$g_1(u)=1$ and  $g_2{(u)}$ any of the functions \eqref{g1g2}, or
combinations of functions $$(g_1{(u)}, g_2{(u)}) \in\left\{
(u^{-2},\tanh^2u),(u^{-2},\coth^2u),(\tanh^2u,\tanh^2u),(\coth^2u,\coth^2u)\right\}.$$
As we shall see, dualization with spectators yields also other cases
of (\ref{sfets}) as duals of the flat metric.

Let us bring the
backgrounds obtained by duality with respect to subgroups generated
by algebras (\ref{S1,x})-(\ref{S3,x}) to the Brinkmann and Rosen
\coor s.
\subsubsection{{Plane-parallel waves as duals with respect to one-dimensional subgroups}}
\begin{enumerate}
\item {Dual obtained with respect to subgroup generated by $S_{1,4}$}

Using the transformation of coordinates
\begin{equation}
s_1={u},\ s_2=-\frac{{u}^2}{2}+{u}
   \,{v}-\frac{{z_3}^2}{2},\ s_3={z_4},\ \ghat_1={u} {z_3},
\end{equation} the symmetric dual tensor %(\ref{whF1,4})
obtained by \dualn{} with respect to subgroup generated by algebra $S_{1,4}$ {(see Tab. \ref{table4})} can be
transformed to the metric in the Brinkmann {coordinates}
\begin{equation}\label{BrinkMetrics14}
   ds^{2}= 2dudv+2\frac{z_3^2}{u^2}du^2+d{z_3}^2+d{z_4}^2.
\end{equation}
{The torsion vanishes and dilaton field satisfying the \eqn{}
\eqref{dilaton_eqn} is $$\phi(u)= c_2+c_1u+2\log(u)$$ that corresponds to the formula for the \tfn{} of the dilaton fields  \cite{delaossa:1992vc}
\begin{equation}
\phi(s,\ghat)=\phi_0(s)-\log(\det M(s,\ghat)).
\end{equation}}
Subsequent
\tfn{} to the Rosen \coor s yields
\begin{equation}\label{RosenMetrics14}
    ds^{2}= 2dudv+dx_1^2+\frac{1}{u^2}\,dx_2^2.
\end{equation}
The latter expression is of the Tseytlin form (\ref{sfets}) with
$g_1(u)=u^2,\ g_2(u)=1,\ q=0$. {\emph{This background was not found\footnote{{However, the case with $g_1(u)=u^2,\ g_2(u)=1,\ q=1$ did appear in \cite{hlapet:cqgrav}.}} by atomic duality in \cite{hlapet:cqgrav}.}}

\item {Dual obtained with respect to subgroup generated by $S_{1,13}$}

The symmetric part of tensor %(\ref{whF1,13})
obtained by \dualn{} with respect to subgroup generated by algebra $S_{1,13}$ (see Tab. \ref{table4}) can be
transformed by
\begin{equation}\ s_1={u},\ s_2=\frac{{z_3}^2}{2 \left({u}^2+1\right)}-\frac{{u}^2
   {z_4}^2}{2 \left({u}^2+1\right)}-\frac{{u}^2}{2}+{u}
   \,{v},\ s_3= {z_3}\frac{\sqrt{{u}^2+1}}{{u}},\ \ghat_1={z_4}\sqrt{{u}^2+1}
\end{equation}to the metric in the Brinkmann coordinates
\begin{equation}\label{BrinkMetrics113}
   ds^{2}= 2dudv-\frac{3 z_3^2+(1-2u^2)z_4^2}{(1+u^2)^2}du^2+d{z_3}^2+d{z_4}^2
.\end{equation} This metric (as well as torsion and dilaton) {already} appeared in {Ref.} \cite{hlapet:cqgrav}
by \dualn{} with respect to algebra\footnote{{The algebras
denoted as $S_{4,n}$ are four-dimensional algebras $S_n$ used in Ref. \cite{hlapet:cqgrav} with
the number $n$ of the subalgebra in Tab. IV, \cite{PWZ}.}} $S_{4,23}$ and $S_{4,25}$. Its Rosen form
\begin{equation}\label{RosenMetrics113}
    ds^{2}= 2dudv+\frac{u^2}{1+u^2}\,dx_1^2+\frac{1}{1+u^2}\,dx_2^2
\end{equation}
is given by (\ref{sfets}) with $g_1(u)=u^2,\ g_2(u)=1,\
q= 1$. {The torsion that follows from the antisymmetric part
is} \begin{equation}
H= \frac{2}{1+u^2}du\wedge dx_1\wedge dx_2.
\end{equation}

\end{enumerate}

\subsubsection{Plane-parallel waves as duals with respect to two-dimensional subgroups}

\begin{enumerate}
\item {Dual obtained with respect to subgroup generated by $S_{2,2}$}

The symmetric dual tensor %(\ref{whF2,2})
obtained by \dualn{} with respect to subgroup generated by algebra $S_{2,2}$ {(see Tab. \ref{table5})} can be
transformed by
\begin{equation}
\ s_1={u},\ s_2=-{u}^2+2 {u} \,{v}-{z_3}^2-{z_4}^2, \ \ghat_1={u}
   {z_3},\ \ghat_2={u} {z_4}
\end{equation}
to the metric in the Brinkmann \coor s
\begin{equation}\label{BrinkMetrics22}
   ds^{2}= 2dudv+2\frac{{z_3}^2+{z_4}^2}{u^2}du^2+d{z_3}^2+d{z_4}^2.
\end{equation}
{The torsion vanishes and solution of the \eqn{}
\eqref{dilaton_eqn} is $$ \phi(u)= c_2+c_1u+4\log(u).$$

Rosen form
of the metric }
\begin{equation}\label{RosenMetrics22}
    ds^{2}= 2dudv+\frac{1}{u^2}\,dx_1^2+\frac{1}{u^2}\,dx_2^2
\end{equation}
is given by (\ref{sfets}) with $g_1(u)=g_2(u)=u^2,\ q=0.$ {\emph{ This
background was not found  by atomic duality.}}

\item {Dual obtained with respect to subgroup generated by $S_{2,8}$}

The symmetric part of dual tensor %(\ref{whF2,8})
obtained by \dualn{} with respect to subgroup generated by algebra $S_{2,8}$ (see Tab. \ref{table5}) can be
transformed by
\begin{equation}
s_1={u},\ s_2=-u^2+2{u}\,{v}-{z_3}^2,\ \ghat_1=
   {u}\,{z_3},\ \ghat_2= {z_4}
\end{equation}
to the metric (\ref{BrinkMetrics14}) and consequently to the Rosen form (\ref{RosenMetrics14}). The torsion vanishes.
\item {Dual obtained with respect to subgroup generated by $S_{2,12}$}

The symmetric dual tensor %(\ref{whF2,12})
obtained by \dualn{} with respect to subgroup generated by algebra $S_{2,12}$ (see Tab. \ref{table5}) can be
transformed by
{
\begin{equation}
s_1={u},\ s_2=\frac{u+1}{2} \left(2\,{v}-\frac{{z_3}^2}{{u}}-{u}\right)-\frac{{z_4}^2}{2},\ \ghat_1={u}{z_3},\ \ghat_2=({u}+1) {z_4}
\end{equation}}
to the metric {in the Brinkmann coordinates}
\begin{equation}\label{BrinkMetrics212}
    2dudv+2\left(\frac{{z_3}^2}{u^2}+\frac{{z_4}^2}{(1+u)^2}\right)du^2+d{z_3}^2+d{z_4}^2.
\end{equation}{The torsion vanishes and solution of the \eqn{}
\eqref{dilaton_eqn} is $$ \phi(u)= c_2+c_1u+2\log(u+u^2).$$
Subsequent \tfn{} to the Rosen \coor s yields}

\begin{equation}\label{RosenMetrics212}
    ds^{2}= 2dudv+\frac{1}{u^2}\,dx_1^2+\frac{1}{(1+u)^2}\,dx_2^2
\end{equation}
given by (\ref{sfets}) with $g_1(u')=(1+u)^2,\ g_2(u)=u^2,\ q=0$.
{\emph{ This background was not found   by atomic duality.}}

\item {Dual obtained with respect to subgroup generated by $S_{2,13}$}

The symmetric part of dual tensor %(\ref{whF2,13})
obtained by \dualn{} with respect to subgroup generated by algebra $S_{2,13}$ (see Tab. \ref{table5}) can be
transformed by
\begin{equation}
s_1={z_4},\ s_2={z_3},\ \ghat_1={v}
   ,\ \ghat_2=-2\,\epsilon\,{u}
\end{equation}
to the metric {in the Brinkmann coordinates}
\begin{equation}\label{BrinkMetrics213}
   ds^{2}= 2dudv-\left({z_3}^2+{z_4}^2\right)du^2+d{z_3}^2+d{z_4}^2.
\end{equation}
 This metric {already} appeared in {Ref.} \cite{hlapet:cqgrav}
 by \dualn{} with respect to algebras $S_{4,17},S_{4,29}$. Its Rosen form
\begin{equation}\label{RosenMetrics213}
    ds^{2}= 2dudv+\cos^2u\,dx_1^2+\sin^2u\,dx_2^2
\end{equation}
is given by (\ref{sfets}) with $g_1(u)=1,\ g_2({u})=\tan^2 u,\
q=-1$. The torsion obtained from the antisymmetric part of dual tensor is
\begin{equation}\label{torsion213}
H=\sin(2u)\,du\wedge dx_1\wedge dx_2. \end{equation}
\item {Dual obtained with respect to subgroup generated by $S_{2,21}$}

The symmetric part of dual tensor %(\ref{whF2,21})
obtained by \dualn{} with respect to subgroup generated by algebra $S_{2,21}$ (see Tab. \ref{table6}) can be
transformed to the metric (\ref{BrinkMetrics213}) by
$$s_1={z_3} \cos \left[\cot \gamma \log (\cosh (u\,\tan \gamma))\right]+{z_4} \sin \left[\cot \gamma \log (\cosh
   (u\,\tan \gamma))\right],$$
   \begin{equation}\label{tfn221}
   s_2={z_4} \cos \left[\cot \gamma \log (\cosh (u\,\tan \gamma))\right]-{z_3} \sin \left[\cot
   \gamma \log (\cosh (u\,\tan \gamma))\right],
   \end{equation}
   $$\ghat_1={v} \cos \gamma,\ \ghat_2=\tanh (u\,\tan \gamma)$$
for $|\ghat_2|<1$ or {changing} $\cosh (u\,\tan \gamma)\, \rightarrow\,\sinh (u\,\tan \gamma),\ \tanh
(u\,\tan \gamma)\,\rightarrow\, \coth (u\,\tan \gamma)$ in (\ref{tfn221}) for $|\ghat_2|>1$.  The torsion obtained from the antisymmetric part of dual tensor is (\ref{torsion213}).

\end{enumerate}

\subsubsection{Plane-parallel waves as duals with respect to three-dimensional subgroups}

\begin{enumerate}

\item {Dual obtained with respect to subgroups generated by $S_{3,22}$, $S_{3,25}$}

{The symmetric part of dual tensor
%(\ref{whF3,22}), (\ref{whF3,23})
obtained by \dualn{} with respect to subgroups generated by algebras $S_{3,22}$ and
$S_{3,25}$ {(see Tab. \ref{table8})} can be transformed by ($\alpha=0$ for $S_{3,25}$)}
$$s_1={u},$$
\begin{equation}\label{tfn322}
\ghat_1=\frac{\alpha^2}{2}u + v+\alpha\,z_3(\log(\cosh u)-1)- \frac{1}{2}
   \tanh u\left[\alpha^2\left((\log (\cosh u)-1)^2+{z_3}^2\right)\right],
\end{equation}
$$\ghat_2=z_3- \alpha \tanh u\log (\cosh u),\ \ghat_3=\tanh u$$
for $|\ghat_3|<1$ to the metric {in the Brinkmann coordinates}
\begin{equation}\label{BrinkMetrics322}
   ds^{2}= 2dudv-2\frac{{z_3}^2}{\cosh^2 u}du^2+d{z_3}^2+d{z_4}^2.
\end{equation}
{Changing} $\cosh u\, \rightarrow\,\sinh u,\ \tanh
u\,\rightarrow\, \coth u$ in (\ref{tfn322}) for $|\ghat_3|>1$, we get
\begin{equation}\label{BrinkMetrics322b}
   ds^{2}= 2dudv+2\frac{{z_3}^2}{\sinh^2 u}du^2+d{z_3}^2+d{z_4}^2.
\end{equation}
The torsions {vanish}. These metrics {already} appeared in {Ref.}
\cite{hlapet:cqgrav} by \dualn{} with respect to algebras
$S_{4,7},S_{4,8}$.
If we neglect the antisymmetric part of dual tensor %(\ref{whF3,22}),
because of vanishing torsion, the Rosen form of
(\ref{BrinkMetrics322}) and (\ref{BrinkMetrics322b}) are given by
(\ref{sfets}) with $g_1(u)=1,\ g_2(u)=\coth^2(u),\ q=0$ and
$g_1(u)=1,\ g_2(u)=\tanh^2(u),\ q=0$.

\item {Dual obtained with respect to subgroups generated by $S_{3,23}$, $S_{3,26}$}

The symmetric part of dual tensor %(\ref{whF3,23})
obtained by \dualn{} with respect to subgroups generated by algebras $S_{3,23}$ and
$S_{3,26}$ (see Tab. \ref{table8}) can be transformed by ($\beta=0$ for $S_{3,26}$)
$$s_1={u}\, e^{\frac{{z_4}}{\alpha}},$$
\begin{equation}\label{tfn323}
\begin{array}{c}
\ghat_1=\text{sgn}\, u\Big[ v+\alpha^2
   \int \frac{\tanh u}{{u}} \,d{u}+\frac{\alpha^2-2 \alpha {z_4}}{2 {u}}+\frac{\beta}{2}\left[
   \beta\, {u}-2 {z_3}   -  {z_3} \log
   \left(\text{sech}^2u\right)\right]\\
   -\frac{1}{8}
   \tanh u\left[
  4 \left(\beta^2-\alpha (\alpha+2 {z_4})+{z_3}^2\right)+
   \beta^2 \log
   \left(\text{sech}^2u\right) \left(\log
   \left(\text{sech}^2u\right)+4\right)\right]\Big],\\
\end{array}
\end{equation}
$$\ghat_2=\frac{1}{2} \beta \tanh u \log \left(\text{sech}^2u\right)+{z_3},\ \ghat_3=\tanh u$$
for $|\ghat_3|<1$  to the metric {in the Brinkmann coordinates}
\begin{equation}\label{BrinkMetrics323}
   ds^{2}= 2dudv-2\left(\frac{{z_3}^2}{\cosh^2 u}-\alpha\,z_4\,\left(\frac{1}{u^2}+\frac{1}{\cosh^2 u}\right)\right)du^2+d{z_3}^2+d{z_4}^2.
\end{equation}Again, {changing} $\text{sech}\,u\, \rightarrow\,\text{csch}\, u,\ \tanh
u\,\rightarrow\, \coth u$ in (\ref{tfn323}) for $|\ghat_3|>1$, we get
\begin{equation}\label{BrinkMetrics323b}
   ds^{2}= 2dudv+2\left(\frac{{z_3}^2}{\sinh^2 u}-\alpha\,z_4\,\left(\frac{1}{u^2}+\frac{1}{\sinh^2 u}\right)\right)du^2+d{z_3}^2+d{z_4}^2.
\end{equation}The torsions {vanish}.

{The linear dependence of metrics on $z_4$ can be eliminated
by
$$ (u,v,z_3,z_4)\mapsto(u,v-a'_1(u)z_4 +a_2(u),z_3,z_4+a_1(u)),$$ with
$$ a_1(u)=\half \int\left(\int L(u) du\right) du, \ a_2(u)=-\half \int\left( a_1(u)L(u)+a'_1(u)^2\right) du$$ where
$ L(u)$ is the coefficient of the linear term in $z_4$. By this subsequent \tfn{} we can bring the metrics (\ref{BrinkMetrics323}) and (\ref{BrinkMetrics323b}) to the forms \eqref{BrinkMetrics322} and \eqref{BrinkMetrics322b}.}
\end{enumerate}

\subsection{Diagonalizable metrics}
{Besides pp-waves, all remaining dual metrics turn out to be of the 1+3 block diagonal form. As a result, they might be in principle diagonalized (see e.g. {Ref.} \cite{grantvic} and references therein). We have managed to do that in most of the cases with the exception of dual metrics with non-vanishing scalar curvature following} from the subalgebras
$$ S_{1,1},\ S_{1,8},\ S_{2,23},\ S_{3,24},\ S_{3,31},\ S_{3,37}.$$
These backgrounds are listed in the Appendix.

\subsubsection{Flat backgrounds}
Backgrounds with flat metric are obtained as duals with respect to
subgroups generated by subalgebras
$$ S_{1,6},\ S_{1,7},$$ $$ S_{2,10},\  S_{2,11},\  S_{2,17},\  S_{2,22},\  S_{2,24},$$ $$S_{3,6},\ S_{3,7},\  S_{3,9},\ S_{3,12},\ S_{3,13},\  S_{3,18},\ S_{3,20}.$$
All {of} the corresponding dual sigma models have vanishing torsion. Duality transformation in these cases represents a mere change of  coordinates. Therefore, we do not discuss these duals further.

\subsubsection{Cases obtainable by atomic duality}

In this {subsection}, we summarize dual backgrounds whose form already
appeared in {Ref.} \cite{hlapet:cqgrav} where the results were obtained
via atomic duality. These duals resemble black-hole-type or cosmological backgrounds reviewed in Ref. \cite{tsey:revExSol} as solutions corresponding to gauged WZW models.

\begin{enumerate}
\item Dual obtained with respect to subgroups generated by  $S_{1,9}, S_{2,14}, S_{3,35}$

{Using the transformation of coordinates
\begin{equation}
s_1=y_4,\ s_2=y_3\sin \frac{y_2}{\alpha},\ s_3=y_3\cos \frac{y_2}{\alpha},\ \hat{g}_1=y_1,
\end{equation}
\begin{equation}
s_1=y_3\cos \frac{y_2}{\alpha},\ s_2=y_3\sin \frac{y_2}{\alpha},\ \hat{g}_1=y_1,\ \hat{g}_2=y_4,
\end{equation}
\begin{equation}
s_1=y_4,\ \ \hat{g}_1=y_1,\ \hat{g}_2=y_3\cos \frac{y_2}{\alpha},\ \hat{g}_3=y_3\sin \frac{y_2}{\alpha}
\end{equation}
for $S_{1,9}$, $S_{2,14}$, and $S_{3,35}$ respectively, the} metric of the dual background acquires the form 
\begin{equation}
ds^{2}= \frac{1}{y_3^2-\alpha ^2}\,d{y_1}^2+\frac{y_3^2
}{\alpha ^2-y_3^2}\,d{y_2}^2+d{y_3}^2+d{y_4}^2,
\end{equation}where $y_3\neq|\alpha|$.
The torsion is
\begin{equation}
H=\frac{\epsilon\, 2 y_3
\alpha}{\left(y_3^2-\alpha^2\right)^2}\,dy_1\wedge dy_2\wedge
dy_3
\end{equation}
where $\alpha >0$, $\epsilon = 1$ for $S_{1,9}, S_{3,35}$ and $\epsilon = -1$ for
$S_{2,14}$.

\item Dual obtained with respect to subgroup generated by $S_{3,29}$

Using the transformation of coordinates
\begin{equation}
s_1=y_2,\ \hat{g}_1=y_3,\ \hat{g}_2=y_1\cosh\frac{y_4}{\alpha},\ \hat{g}_3=y_1\sinh\frac{y_4}{\alpha}
\end{equation}
for $S_{3,29}$, the metric of the dual background acquires the form
\begin{equation}
ds^{2}= -d{y_1}^2+d{y_2}^2+\frac{1}{y_1^2+\alpha
^2}\,d{y_3}^2+\frac{y_1^2}{y_1^2+\alpha ^2}\,d{y_4}^2,
\end{equation}
while the torsion is
\begin{equation}
H=\frac{ 2 y_1 \alpha}{\left(y_1^2+\alpha^2\right)^2}\,dy_1\wedge
dy_3\wedge dy_4
\end{equation}
where $\alpha >0$.

\item Dual obtained with respect to subgroups generated by $S_{1,10}, S_{2,15}, S_{3,36}$

{Using the transformation of coordinates
\begin{equation}
s_1=y_1,\ s_2=y_2^2,\ s_3=y_3,\ \hat{g}_1=y_4,\quad s_2> 0,
\end{equation}
\begin{equation}
s_1=y_2^2,\ s_2=y_3,\ \hat{g}_1=y_4,\ \hat{g}_2=y_1,\quad s_1> 0,
\end{equation}
\begin{equation}
s_1=y_1,\ \hat{g}_1=y_4,\ \hat{g}_2=y_2\cos \frac{y_3}{\alpha},\ \hat{g}_3=y_2\sin \frac{y_3}{\alpha}
\end{equation}
for $S_{1,10}$, $S_{2,15}$, and $S_{3,36}$ respectively, the} metric of the dual background acquires the form
\begin{equation}
ds^{2}= -d{y_1}^2+d{y_2}^2+\frac{y_2^2}{y_2^2+\alpha
^2}\,d{y_3}^2+\frac{1}{y_2^2+\alpha ^2}\,d{y_4}^2,
\end{equation}
while the torsion is
\begin{equation}
H=\frac{\epsilon\, 2 y_2
\alpha}{\left(y_2^2+\alpha^2\right)^2}\,dy_2\wedge dy_3\wedge dy_4
\end{equation}
where $\alpha \neq 0$, $\epsilon = 1$ for $S_{1,10},S_{3,36}$ and $\epsilon = -1$ for
$S_{2,15}$.
\end{enumerate}

\subsubsection{Newly obtained diagonal metrics}
In this {subsection}, we {list} newly obtained dual backgrounds for which
we {are} able to explicitly find coordinate systems in which dual metrics become diagonal.

\begin{enumerate}
\item Dual obtained with respect to subgroups generated by $S_{1,2}, S_{2,4}, S_{2,5}, S_{3,2}$

The dual background is torsionless. {Using the transformation of coordinates
\begin{equation}
s_1=y_1,\ s_2=y_2,\ s_3=4y_4,\ \hat{g}_1=2y_3,
\end{equation}
\begin{equation}
s_1=y_1,\ s_2=4y_4,\ \hat{g}_1=2y_3,\ \hat{g}_2=y_2,
\end{equation}
\begin{equation}
s_1=y_2,\ s_2=4y_4,\ \hat{g}_1=2y_3,\ \hat{g}_2=y_1,
\end{equation}
\begin{equation}
s_1=4 y_4,\ \hat{g}_1=2y_3,\ \hat{g}_2=y_1,\ \hat{g}_3=y_2,
\end{equation}
for $S_{1,2}$, $S_{2,4}$, $S_{2,5}$, and $S_{3,2}$ respectively, the} metric of the dual background acquires the form
\begin{equation}\label{Kmodel}
ds^{2}=
-d{y_1}^2+d{y_2}^2+\frac{1}{y_4}d{y_3}^2+\frac{1}{y_4}d{y_4}^2
\end{equation}
where $y_4> 0$.
This background might be considered a special case of K-models discussed in Ref. \cite{hortseyt} as generalization of plane wave  \soln.

\item Dual obtained with respect to subgroups generated by $S_{1,3}$, $S_{2,6}, S_{3,3}$

The dual background is torsionless. {Using the transformation of
coordinates
\begin{equation}
s_1=y_4,\ s_2=y_3,\ s_3=-4y_1,\ \hat{g}_1=2y_2,
\end{equation}
\begin{equation}
s_1=4y_1,\ s_2=y_3,\ \hat{g}_1=2y_2,\ \hat{g}_2=y_4,
\end{equation}}
\begin{equation}
s_1=4y_1,\ \hat{g}_1=2{y_2},\ \hat{g}_2=y_3,\ \hat{g}_3=y_4
\end{equation}
{for $S_{1,3}$, $S_{2,6}$, and $S_{3,3}$ respectively, the} metric of the dual background acquires the form 
\begin{equation}
ds^{2}=
-\frac{1}{y_1}d{y_1}^2+\frac{1}{y_1}d{y_2}^2+d{y_3}^2+d{y_4}^2,
\end{equation}where $y_1\neq 0$.

\item Dual obtained with respect to subgroup generated by $S_{2,1}$

The dual background is torsionless. Using the transformation of
coordinates
\begin{equation}
s_1=4y_1,\ s_2=4y_4,\ \hat{g}_1=2y_3,\ \hat{g}_2=2y_2
\end{equation}
for $S_{2,1}$, the metric of the dual background acquires the form
\begin{equation}
ds^{2}=
-\frac{1}{y_1}d{y_1}^2+\frac{1}{y_1}d{y_2}^2+\frac{1}{y_4}d{y_3}^2+\frac{1}{y_4}d{y_4}^2
\end{equation}
where $y_4> 0,\ y_1\neq0$.

\item Dual obtained with respect to subgroups generated by $S_{1,12}, S_{2,19}$

{Using the transformation of coordinates
\begin{equation}
s_1=y_3,\ s_2=4y_4,\ s_3=12y_1,\ \hat{g}_1=2y_2,
\end{equation}
\begin{equation}
s_1=4y_4,\ s_2=12y_1,\ \hat{g}_1=2y_2,\ \hat{g}_2=y_3
\end{equation}
for $S_{1,12}$ and $S_{2,19}$ respectively, the} metric of the dual background acquires the form
\begin{equation}\label{Fmodel}
ds^{2}=
-\frac{1}{y_4}d{y_1}^2+\frac{1}{y_4}d{y_2}^2+d{y_3}^2+d{y_4}^2,
\end{equation}
where $y_4\neq 0$. The torsion is
\begin{equation}
H=\frac{1}{y_4^2}\,dy_1\wedge dy_2\wedge dy_4.
\end{equation}
This background belongs to the class of F-models discussed in Ref. \cite{hortseyt} in context of exact string \soln s.

\item Dual obtained with respect to subgroups generated by $S_{1,11}, S_{2,16}$

{Using the transformation of coordinates
\begin{equation}
s_1=y_4,\ s_2=2\alpha \left(y_2-y_3+\frac{\alpha}{2}\right),\
s_3=y_3,\ \hat{g}_1=\sqrt{2\alpha}\,y_1,
\end{equation}
\begin{equation}
s_1=2\alpha \left(y_2-y_3+\frac{\alpha}{2}\right),\ s_2=y_3,\ \hat{g}_1=\sqrt{2\alpha}\,y_1,\ \hat{g}_2=y_4
\end{equation}
for $S_{1,11}$ and $S_{2,16}$ respectively, the} metric of the dual background acquires the form
\begin{equation}
ds^{2}=
-\frac{1}{y_2-y_3}d{y_1}^2+\frac{\alpha}{2y_2-2y_3}d{y_2}^2+d{y_3}^2+d{y_4}^2,
\end{equation}
while the torsion is
\begin{equation}
H=-\frac{\sqrt{\alpha}}{\sqrt{2}\left(y_2-y_3\right)^2}\,dy_1\wedge
dy_2\wedge dy_3
\end{equation}
where $\alpha>0,\ y_2\neq y_3$.

\item Dual obtained with respect to subgroups generated by $S_{2,20}, S_{3,10}$

{Using the transformation of coordinates
\begin{equation}
s_1=y_1,\ s_2=y_4,\ \hat{g}_1=y_2,\ \hat{g}_2=y_3,
\end{equation}
\begin{equation}
s_1=y_1,\ \hat{g}_1=y_2,\ \hat{g}_2=y_3,\ \hat{g}_3=y_4
\end{equation}
for $S_{2,20}$ and $S_{3,10}$ respectively, the metric} of the dual background acquires the form
\begin{equation}
ds^{2}=
-\frac{1}{4y_1}d{y_1}^2+\frac{\sigma y_1}{\sigma y_1^2+y_3^2}d{y_2}^2+\frac{y_1}{\sigma y_1^2+y_3^2}d{y_3}^2+d{y_4}^2,
\end{equation}
while the torsion is
\begin{equation}
H=\frac{2 \sigma  y_1 y_3}{\left(\sigma  y_1^2+y_3^2\right)^2}\,dy_1\wedge
dy_2\wedge dy_3
\end{equation}
where $\sigma = \text{sgn}\,y_1,\ y_1\neq 0$.

\item Dual obtained with respect to subgroup generated by $S_{3,11}$

The dual background is torsionless. Using the transformation of
coordinates
\begin{equation}
s_1=4y_4,\ \hat{g}_1=\frac{1}{2}(y_2-y_1),\ \hat{g}_2=y_2+y_1,\
\hat{g}_3=2y_3
\end{equation}
for $S_{3,11}$, the metric of the dual background acquires the form
\begin{equation}
ds^{2}=
-\frac{1}{1-(y_1+y_2)^2}d{y_1}^2+\frac{1}{1-(y_1+y_2)^2}d{y_2}^2+\frac{1}{y_4}
d{y_3}^2+\frac{1}{y_4} d{y_4}^2,
\end{equation}where $y_4>0$.

\item Dual obtained with respect to subgroup generated by $S_{3,27}$

Using the transformation of coordinates
\begin{equation}
s_1=y_2^2, \  \hat{g}_1= y_4\cos\gamma,\
\hat{g}_2=y_1\cot\gamma\cosh(y_3\tan\gamma),\
\hat{g}_3=y_1\cot\gamma\sinh(y_3\tan\gamma)
\end{equation}
for $S_{3,27}$, the metric of the dual background acquires the form
\begin{equation}
ds^2=-(\cot\,\gamma)^2 d{y_1}^2+d{y_2}^2+\frac{y_1^2y_2^2}{y_1^2+y_2^2}
d{y_3}^2+\frac{1}{y_1^2+y_2^2}d{y_4}^2,
\end{equation}
where $0<\gamma<\pi,\ \gamma\neq\frac{\pi}{2},\ y_2\neq 0$, while the torsion is
\begin{equation}
H=-\frac{2y_1y_2^2}{y_1^2+y_2^2}\,dy_1\wedge dy_3\wedge dy_4 +
  \frac{2y_1^2y_2}{y_1^2+y_2^2}\,dy_2\wedge dy_3\wedge dy_4.
\end{equation}
\end{enumerate}

\section{Conclusion} Together with paper \cite{hlapet:cqgrav}, we
have classified backgrounds obtainable as (non-)Abelian T-duals of
the flat Lorentzian metric in four dimensions with respect to subgroups listed in \cite{PWZ}.

{We have circumvented the tedious calculation of the adapted \coor
s by deriving formulas (\ref{Exi blok}), (\ref{Exicomps}) for the spectator dependent matrix $E_\xi$
necessary for construction of dualizable backgrounds.} All the dual
backgrounds are pp-waves or diagonalizable metrics, both with and
without torsion.

{By dualization procedure with spectators, we have found several
cases that already appeared in atomic duality \cite{hlapet:cqgrav}
but we have also obtained new  pp-wave backgrounds as T-duals of the
flat metric, {namely the following metrics and dilatons.}}\begin{enumerate}
    \item{ \begin{equation}\label{BrinkMetrics14bis}
 ds^{2}=
2dudv+2\frac{z_3^2}{u^2}du^2+d{z_3}^2+d{z_4}^2,
\end{equation}
$$ \phi(u)= c_2+c_1u+2\log(u)$$obtained by dualization with respect to the groups generated by
$S_{1,4}$ and $S_{2,8}$,
    \item \begin{equation}\label{BrinkMetrics22bis}
ds^{2}= 2dudv+2\frac{{z_3}^2+{z_4}^2}{u^2}du^2+d{z_3}^2+d{z_4}^2,
\end{equation}$$ \phi(u)=c_2+c_1u+ 4\log(u)$$obtained by dualization with respect to the group generated by
$S_{2,2}$ and
    \item \begin{equation}\label{BrinkMetrics212bis}
     ds^{2}=
2dudv+2\left(\frac{{z_3}^2}{u^2}+\frac{{z_4}^2}{(1+u)^2}\right)du^2+d{z_3}^2+d{z_4}^2,
\end{equation}
$$ \phi(u)=c_2+c_1u+ 2\log(u+u^2)$$obtained by dualization with respect to the group generated by
$S_{2,12}$.}
\end{enumerate}
{Their corresponding parameters  $ (g_1(u'), g_2(u),q) $ in the
Tseytlin form (\ref{sfets}) are
$$(u^2,1,0),\ (u^2,u^2,0),\ ((1+u)^2,u^2,0).$$
Singularities of the metrics \eqref{BrinkMetrics14bis}-\eqref{BrinkMetrics212bis} %can be interpreted 
are physical in the sense of \cite{HorSteif,pandazajic}, namely that they are the points of spacetime where tidal forces diverge and without them the manifold is geodesically incomplete.}

The metric (\ref{BrinkMetrics22bis}) is a special case of solvable
backgrounds investigated in {Refs. \cite{papa,hlapet:ijmp}}. The
fact that this metric  is T-dual of the flat metric explains its
solvability.
%(e.g. by the method of solution described in \cite{hlapet:cqgrav}).
On the other hand, this is the simple case where the Bessel functions
appearing in {Ref.} \cite{papa} pass to combinations of trigonometric
{functions}.

{List of all obtained pp-wave backgrounds is given in {Tab.}
\ref{table11} where we point out that some of them appeared in \cite{hlapet:cqgrav} as well.} 
Besides pp-waves, we have also obtained many diagonal or
diagonalizable metrics. List of metrics that we were able to diagonalize is given in {Tab.} \ref{table13}. Some of them remind physically interesting and investigated backgrounds -- K-model \eqref{Kmodel}, F-model \eqref{Fmodel}, for others we do not see their physical interpretation.

One might be tempted to consider non-Abelian T-duality with spectators {using  isometry group $G$ to be equivalent to the case of atomic duality with
respect to four-dimensional groups  $G\otimes A$ where $A$ is an Abelian group}. In
many cases, this is true but not in general. E.g., the dual
(\ref{BrinkMetrics322}) obtained from the group given by $S_{3,22}$
and one spectator is the same as the dual obtained in {Ref.}
\cite{hlapet:cqgrav} from the group given by
$S_{4,7}=S_{3,22}\oplus (P_2)$. {Similarly, dual tensors
obtained by dualization with respect to groups generated by
$S_{3,25},\ S_{3,29},\ S_{3,35}$, and $S_{3,36}$ with one spectator agree with dual tensors following from
$S_{4,8}=S_{3,25}\oplus (P_2)$, $S_{4,11}=S_{3,29}\oplus (P_1)$,
$S_{4,18}=S_{3,35}\oplus (P_3)$, and $S_{4,19}=S_{3,36}\oplus (P_0)$, respectively.}

On the other hand, there is a dual tensor obtained from the group
given by $S_{4,17}=S_{3,34}\oplus (P_0-P_3)$, but {a dual tensor} with
respect to the group given by $S_{3,34}$ and one spectator does
not exist as condition (\ref{condsAB}) is not satisfied. Situation
with more than one spectator is even more complicated. {There
are only two four-dimensional subgroups of the form $G\otimes A$
where $A$ has dimension 2 or 3. Nevertheless, the sets of duals with respect to
two- or one-dimensional subgroups with two or three
spectators respectively are much richer.}

\begin{table}
\begin{center}
\footnotesize {\renewcommand{\arraystretch}{1.6}
\begin{tabular}{|c | c | c | c | }
\hline %\multirow{2}{*}
          Metric in Brinkmann coordinates & \multirow{2}{*}{Torsions} &  \multirow{2}{*}{Generated by} \\
         Metric in Rosen coordinates
      & &  \\
\hline \hline $\begin{array}{c}
   2dudv+2\frac{z_3^2}{u^2}du^2+d{z_3}^2+d{z_4}^2\\
   2dudv+dx_1^2+\frac{1}{u^2}\,dx_2^2
 \end{array}$  &  $\begin{array}{c}
   0\\
   0
 \end{array}$  & $S_{1,4}, S_{2,8}$\\
\hline $\begin{array}{c}
   2dudv-\frac{3
z_3^2+(1-2 u^2)z_4^2}{(1+u^2)^2}du^2+d{z_3}^2+d{z_4}^2\\
   2dudv+\frac{u^2}{1+u^2}\,dx_1^2+\frac{1}{1+u^2}\,dx_2^2
 \end{array}$  &  $\begin{array}{c}
   \frac{2}{1+u^2}du\wedge dz_3\wedge dz_4 \\
   -\frac{2}{1+u^2}du\wedge dx_1\wedge dx_2
 \end{array}$  & $\begin{array}{c} S_{1,13},\\{S_{4,23},S_{4,25}}\end{array}$
 \\
\hline $\begin{array}{c}
   2dudv+2\frac{{z_3}^2+{z_4}^2}{u^2}du^2+d{z_3}^2+d{z_4}^2\\
   2dudv+\frac{1}{u^2}\,dx_1^2+\frac{1}{u^2}\,dx_2^2
 \end{array}$  &  $\begin{array}{c}
   0\\
   0
 \end{array}$  & $S_{2,2}$\\
\hline $\begin{array}{c}
   2dudv+2\left(\frac{{z_3}^2}{u^2}+\frac{{z_4}^2}{(1+u)^2}\right)du^2+d{z_3}^2+d{z_4}^2\\
    2dudv+\frac{1}{u^2}\,dx_1^2+\frac{1}{(1+u)^2}\,dx_2^2
 \end{array}$  &  $\begin{array}{c}
   0\\
   0
 \end{array}$  & $S_{2,12}$\\
\hline $\begin{array}{c}
   2dudv-\left({z_3}^2+{z_4}^2\right)du^2+d{z_3}^2+d{z_4}^2\\
    2dudv+\cos^2u\,dx_1^2+\sin^2u\,dx_2^2 \end{array}$  &  $\begin{array}{c}
   {2}\,du\wedge dz_3\wedge dz_4\\
   \sin 2u\,du\wedge dx_1\wedge dx_2
 \end{array}$  &  $\begin{array}{c} S_{2,13},S_{2,21},\\{S_{4,17}}\end{array}$\\
\hline $\begin{array}{c}
    2dudv-2\frac{{z_3}^2}{\cosh^2 u}du^2+d{z_3}^2+d{z_4}^2\\
      2dudv+\tanh^2 u\,dx_1^2+u^2\,dx_2^2\end{array}$  & $\begin{array}{c}
  0\\
   0
 \end{array}$  &$\begin{array}{c} S_{3,22},S_{3,23},S_{3,25},S_{3,26}\\{S_{4,7},S_{4,8},S_{4,27}}\end{array}$\\ \hline
$\begin{array}{c}
    2dudv+2\frac{{z_3}^2}{\sinh^2 u}du^2+d{z_3}^2+d{z_4}^2\\
      2dudv+\coth^2 u\,dx_1^2+u^2\,dx_2^2\end{array}$  & $\begin{array}{c}
  0\\
   0
 \end{array}$  &$\begin{array}{c} S_{3,22},S_{3,23},S_{3,25},S_{3,26}\\{S_{4,7},S_{4,8},S_{4,27}}\end{array}$\\
\hline\end{tabular} \normalsize \caption { List of pp-waves
obtainable by {(non)-Abelian}T-duality %with spectators
\label{table11}} }
\end{center}
\end{table}

\begin{table}
\begin{center}
\footnotesize {\renewcommand{\arraystretch}{1.6}
\begin{tabular}{|c | c | c | c | }
\hline
  Diagonal Metric  &  Torsion & Generated by \\
\hline

\hline $
-d{y_1}^2+d{y_2}^2+\frac{1}{y_4}d{y_3}^2+\frac{1}{y_4}d{y_4}^2 $&
  $0 $&
  $S_{1,2},S_{2,4},S_{2,5},S_{3,2}$ \\

\hline
$-\frac{1}{y_1}d{y_1}^2+\frac{1}{y_1}d{y_2}^2+d{y_3}^2+d{y_4}^2 $&
  $0 $&
  ${S_{1,3}},S_{2,6},S_{3,3}$ \\

\hline $ \frac{1}{y_3^2-\alpha ^2}\,d{y_1}^2+\frac{ y_3^2 }{\alpha
^2-y_3^2}\,d{y_2}^2+d{y_3}^2+d{y_4}^2$ &

$\frac{\epsilon\, 2 y_3
\alpha}{\left(y_3^2-\alpha^2\right)^2}\,dy_1\wedge dy_2\wedge
dy_3$ & $S_{1,9}, S_{2,14}, S_{3,35}, {S_{4,18}}$\\

\hline $ -d{y_1}^2+d{y_2}^2+\frac{1}{y_1^2+\alpha
^2}\,d{y_3}^2+\frac{y_1^2}{y_1^2+\alpha ^2}\,d{y_4}^2 $&
  $ \frac{ 2 y_1 \alpha}{\left(y_1^2+\alpha^2\right)^2}\,dy_1\wedge
dy_3\wedge dy_4$&
  $S_{3,29}, {S_{4,11}}$ \\

\hline $ -d{y_1}^2+d{y_2}^2+\frac{y_2^2}{y_2^2+\alpha
^2}\,d{y_3}^2+\frac{1}{y_2^2+\alpha ^2}\,d{y_4}^2 $&
  $\frac{\epsilon\, 2 y_2
\alpha}{\left(y_2^2+\alpha^2\right)^2}\,dy_2\wedge dy_3\wedge dy_4
$&
  $S_{1,10},S_{2,15},S_{3,36}, {S_{4,19}}$ \\

\hline
$-\frac{1}{y_1}d{y_1}^2+\frac{1}{y_1}d{y_2}^2+\frac{1}{y_4}d{y_3}^2+\frac{1}{y_4}d{y_4}^2
 $&
  $0$&
  $S_{2,1}$ \\

\hline
$-\frac{1}{y_4}d{y_1}^2+\frac{1}{y_4}d{y_2}^2+d{y_3}^2+d{y_4}^2
 $&
  $ \frac{1}{y_4^2}\,dy_1\wedge dy_2\wedge dy_4$&
  $S_{1,12},S_{2,19}$ \\

\hline
$-\frac{1}{y_2-y_3}d{y_1}^2+\frac{\alpha}{2y_2-2y_3}d{y_2}^2+d{y_3}^2+d{y_4}^2
 $&
  $-\frac{\sqrt{\alpha}}{\sqrt{2}\left(y_2-y_3\right)^2}\,dy_1\wedge
dy_2\wedge dy_3 $&
  $S_{1,11},S_{2,16}$ \\

\hline
$-\frac{1}{4y_1}d{y_1}^2+\frac{|y_1|}{y_1|y_1|+y_3^2}d{y_2}^2+\frac{y_1}{y_1|y_1|+y_3^2}
d{y_3}^2+d{y_4}^2
 $&
  $\frac{2\epsilon\, |y_1|y_3
}{y_1|y_1|+y_3^2}\,dy_1\wedge dy_2\wedge dy_3$ &
  $S_{2,20},S_{3,10}$ \\

\hline $\frac{1}{1-(y_1+y_2)^2}(-d{y_1}^2+d{y_2}^2)+\frac{1}{y_4}
d{y_3}^2+\frac{1}{y_4} d{y_4}^2
 $&
  $0 $&
  $S_{3,11}$ \\

\hline $-(\cot\,\gamma)^2
d{y_1}^2+d{y_2}^2+\frac{y_1^2y_2^2}{y_1^2+y_2^2}
d{y_3}^2+\frac{1}{y_1^2+y_2^2}d{y_4}^2
 $&
  $\begin{array}{c} -\frac{2y_1y_2^2}{y_1^2+y_2^2}\,dy_1\wedge dy_3\wedge dy_4 +\\
  \frac{2y_1^2y_2}{y_1^2+y_2^2}\,dy_2\wedge dy_3\wedge dy_4\end{array} $&
  $S_{3,27}$ \\

\hline
\end{tabular}
\normalsize \caption { List of diagonal metrics  obtainable by
T-duality with spectators \label{table13}}
}\end{center}
\end{table}

\section*{{Acknowledgment}}
{This work was supported by the Grant Agency of the Czech Technical University
in Prague, grant No. SGS16/239/OHK4/3T/14.}

%%%%%%%%%%%%%%%%%%%%%%%%%%%%%%%%%%%%%%%%%%%%%%%%%%%%%%%%%%%%%%%%%%%%%%%%%%%%%%%
%% Appendix
%%%%%%%%%%%%%%%%%%%%%%%%%%%%%%%%%%%%%%%%%%%%%%%%%%%%%%%%%%%%%%%%%%%%%%%%%%%%%%%
%\newpage
\section*{Appendix: Duals of flat metric}

Below, we summarize results of dualization of the flat metric by
formulas {(\ref{Exi blok})-(\ref{matsAB}), (\ref{Pihat2}), and (\ref{Fghat2}) using special choices of
functions  $\xi^\mu(s^\delta)$}. Restrictions on spectators
{$s^\delta$} follow from the transitivity of subgroup action, choice
of \fn s $\Phi^\delta$, and conditions (\ref{submanifolds in
e}), (\ref{condsAB}) of dualizability. {In the values of spectators where these \cond s are not satisfied singularities of the dual backgrounds may appear. }

For better readability, we have replaced superscripts of spectators
$s^\delta$ by subscripts $s_\delta$ in the tables. {The
comment not dualizable means that there are no \fn s $\xi^\mu$ that
satisfy the \cond s (\ref{submanifolds in e}), (\ref{condsAB}). In principle diagonalizable means that there are diagonalizing \coor s  \cite{grantvic} but we are not able to find them.}

\begin{table}
\begin{center}
\scriptsize {\renewcommand{\arraystretch}{1.4}
\begin{tabular}{|c || c | c | c |}
\hline
%Subgroup
 & Choice of $\xi^{0}, \xi^{1}, \xi^{2}, \xi^{3}$ & Dual background $\wh F_\xi
(s,\ghat)$ & Comments\\
\hline \hline $S_{1,1}$ & $\begin{array}{c} \frac{1}{2}(1-s_1),\,
s_2,\, s_3,\, \frac{1}{2} (1+s_1), \\ {\cos}(\gamma )^2 \left(s_2^2+s_3^2\right)-s_1\,
{\sin}(\gamma )^2\neq 0  \end{array}$ & see $\wh
F_{1,1}(s,\ghat)$ below
 & {\propp{in principle}{diagonalizable metric}{with torsion}} \\
\hline
$S_{1,2}$ &  $\begin{array}{c} s_1,\, \sqrt{s_3},\, 0,\, s_2, \\ s_3>0 \end{array}$ & $\text{diag}\left(-1, 1, \frac{1}{4 s_3}, \frac{1}{s_3}\right)$ & {\prop{diagonal metric}{torsionless}} \\
\hline $S_{1,3}$ & $\begin{array}{c} \frac{1}{2} (-1+s_3),\, s_1, s_2,\, \frac{1}{2} (1+s_3), \\ s_3\neq 0 \end{array}$ & $\left(
\begin{array}{cccc}
 1 & 0 & 0 & 0 \\
 0 & 1 & 0 & 0 \\
 0 & 0 & \frac{1}{4 s_3} & \frac{1}{2 s_3} \\
 0 & 0 & -\frac{1}{2 s_3} & -\frac{1}{s_3}
\end{array}
\right)$ & {\prop{diagonal metric}{torsionless}} \\
\hline $S_{1,4}$ & $\begin{array}{c} -\frac{s_2}{s_1}, \,  0,\, s_3,\,
 s_1+\frac{s_2}{s_1}, \\ s_1\neq 0\end{array}$ & $\left(
\begin{array}{cccc}
 1-\frac{2 s_2}{s_1^2} & \frac{1}{s_1} & 0 & 0 \\
 \frac{1}{s_1} & 0 & 0 & 0 \\
 0 & 0 & 1 & 0 \\
 0 & 0 & 0 & \frac{1}{s_1^2}
\end{array}
\right)$ & $\begin{array}{c} \text{pp-wave metric}, \\ \text{torsionless} \end{array}$\\
\hline
$S_{1,5}$ & & & not dualizable \\
\hline
$S_{1,6}$ & $s_1, \,  s_2, \,  s_3, \,  0$ & $\text{diag}\left(-1, 1, 1, 1\right)$ & $\begin{array}{c} \text{flat metric}, \\ \text{torsionless} \end{array}$\\
\hline
$S_{1,7}$ & $0, \,  s_1, \,  s_2, \,  s_3$ & $\text{diag}\left(1, 1, 1, -1\right)$ & $\begin{array}{c} \text{flat metric}, \\ \text{torsionless} \end{array}$ \\
\hline $S_{1,8}$ & $\begin{array}{c} 0, \,  s_2, \,  s_3, \,  s_1, \\ s_2^2+s_3^2\neq 0 \end{array}$ & $\left(
\begin{array}{cccc}
 \frac{s_2^2+s_3^2-\epsilon ^2}{s_2^2+s_3^2} & \frac{-\epsilon  s_3}{s_2^2+s_3^2} & \frac{\epsilon  s_2}{s_2^2+s_3^2} & \frac{-\epsilon }{s_2^2+s_3^2} \\
 \frac{-\epsilon  s_3}{s_2^2+s_3^2} & \frac{s_2^2}{s_2^2+s_3^2} & \frac{s_2 s_3}{s_2^2+s_3^2} & \frac{-s_3}{s_2^2+s_3^2} \\
 \frac{\epsilon  s_2}{s_2^2+s_3^2} & \frac{s_2 s_3}{s_2^2+s_3^2} & \frac{s_3^2}{s_2^2+s_3^2} & \frac{s_2}{s_2^2+s_3^2} \\
 \frac{\epsilon }{s_2^2+s_3^2} & \frac{s_3}{s_2^2+s_3^2} & \frac{-s_2}{s_2^2+s_3^2} & \frac{1}{s_2^2+s_3^2}
\end{array}
\right)$ & {\propp{in principle}{diagonalizable metric}{with torsion}}\\
\hline $S_{1,9}$ & $\begin{array}{c} 0, \,  s_3, \,  s_2, \,  s_1,\\
s_2^2+s_3^2\neq \alpha^2 \end{array}$ & $\left(
\begin{array}{cccc}
 1 & 0 & 0 & 0 \\
 0 & \frac{-\alpha ^2+s_2^2}{-\alpha ^2+s_2^2+s_3^2} & \frac{s_2 s_3}{-\alpha ^2+s_2^2+s_3^2} & \frac{s_3}{-\alpha ^2+s_2^2+s_3^2} \\
 0 & \frac{s_2 s_3}{-\alpha ^2+s_2^2+s_3^2} & \frac{-\alpha ^2+s_3^2}{-\alpha ^2+s_2^2+s_3^2} & \frac{-s_2}{-\alpha ^2+s_2^2+s_3^2} \\
 0 & \frac{-s_3}{-\alpha ^2+s_2^2+s_3^2} & \frac{s_2}{-\alpha ^2+s_2^2+s_3^2} & \frac{1}{-\alpha ^2+s_2^2+s_3^2}
\end{array}
\right)$ & {\prop{diagonalizable metric}{with torsion}}\\
\hline $S_{1,10}$ & $\begin{array}{c} s_1, \, 0, \, \sqrt{s_2}, \,
s_3,\\ s_2> 0 \end{array}$ & $\left(
\begin{array}{cccc}
 -1 & 0 & 0 & 0 \\
 0 & \frac{1}{4 s_2} & 0 & 0 \\
 0 & 0 & \frac{s_2}{\alpha ^2+s_2} & \frac{-\alpha }{\alpha ^2+s_2} \\
 0 & 0 & \frac{\alpha }{\alpha ^2+s_2} & \frac{1}{\alpha ^2+s_2}
\end{array}
\right)$ & {\prop{diagonal metric}{with torsion}}\\
\hline $S_{1,11}$ & $\begin{array}{c} \frac{1}{2} (1-s_2),\, s_3,\,
s_1,\, -\frac{1}{2} (1+s_2), \\ s_2\neq \alpha^2 \end{array}$ &
$\left(
\begin{array}{cccc}
 1 & 0 & 0 & 0 \\
 0 & \frac{-1}{4 \left(\alpha ^2-s_2\right)} & \frac{-\alpha }{2 \left(\alpha ^2-s_2\right)} & \frac{-1}{2 \left(\alpha ^2-s_2\right)} \\
 0 & \frac{-\alpha }{2 \left(\alpha ^2-s_2\right)} & \frac{-s_2}{\alpha ^2-s_2} & \frac{-\alpha }{\alpha ^2-s_2} \\
 0 & \frac{1}{2 \alpha ^2-2 s_2} & \frac{\alpha }{\alpha ^2-s_2} & \frac{1}{\alpha ^2-s_2}
\end{array}
\right)$ & {\prop{diagonalizable metric}{with torsion}}\\
\hline $S_{1,12}$ & $\begin{array}{c}-\frac{s_3}{12}, \,
\frac{s_2}{4}, \, s_1, \, \frac{s_3}{12}, \\ s_2\neq 0\end{array}$ &
$\left(
\begin{array}{cccc}
 1 & 0 & 0 & 0 \\
 0 & \frac{1}{16} & 0 & 0 \\
 0 & 0 & -\frac{1}{36 s_2} & -\frac{1}{6 s_2} \\
 0 & 0 & \frac{1}{6 s_2} & \frac{1}{s_2}
\end{array}
\right)$ & {\prop{diagonal metric}{with torsion}}\\
\hline $S_{1,13}$ & $\begin{array}{c} -\frac{s_2}{s_1}, \,  0,\,
s_3,\, s_1+\frac{s_2}{s_1}\end{array}$ & $\left(
\begin{array}{cccc}
 1-\frac{2 s_2}{s_1^2} & \frac{1}{s_1} & 0 & 0 \\
 \frac{1}{s_1} & 0 & 0 & 0 \\
 0 & 0 & \frac{s_1^2}{\epsilon ^2+s_1^2} & \frac{-\epsilon }{\epsilon ^2+s_1^2} \\
 0 & 0 & \frac{\epsilon }{\epsilon ^2+s_1^2} & \frac{1}{\epsilon ^2+s_1^2}
\end{array}
\right)$ & {\prop{pp-wave metric}{with torsion}}\\
\hline
\end{tabular}\normalsize\caption{Dual backgrounds for one-dimensional isometry
groups \label{table4}}
}
\end{center}
\normalsize
\end{table}
\newpage
\footnotesize
$$\wh F_{1,1} (s,\ghat)=f(s,\gamma)\left(
\begin{array}{cccc}
 -\frac{1}{4} \sin ^2(\gamma) & \frac{1}{2} {s_3}
   \cos (\gamma) \sin (\gamma) & -\frac{1}{2} {s_2} \cos
   (\gamma) \sin (\gamma) & \frac{\sin (\gamma)}{2} \\
 \frac{1}{2} {s_3} \cos (\gamma) \sin (\gamma) &
   {s_2}^2 \cos ^2(\gamma)-{s_1} \sin ^2(\gamma) &
   {s_2} {s_3} \cos ^2(\gamma) & -{s_3} \cos
   (\gamma) \\
 -\frac{1}{2} {s_2} \cos (\gamma) \sin (\gamma) &
   {s_2} {s_3} \cos ^2(\gamma) & {s_3}^2 \cos
   ^2(\gamma)-{s_1} \sin ^2(\gamma) & {s_2} \cos (\gamma)
   \\
 -\frac{\sin (\gamma)}{2} & {s_3} \cos (\gamma) &
   -{s_2} \cos (\gamma) & 1 \\
\end{array}
\right),$$\normalsize where \footnotesize$$ f(s,\gamma)=
\frac{1}{{\cos}(\gamma )^2 \left(s_2^2+s_3^2\right)-s_1\,
{\sin}(\gamma )^2}.$$
\normalsize

%input{DualsDim2}
\begin{table}
\begin{center}
\scriptsize {\renewcommand{\arraystretch}{1.4}
\begin{tabular}{|c || c | c | c |}
\hline
%Subgroup
 & Choice of $\xi^{0}, \,\xi^{1}, \,\xi^{2}, \,\xi^{3}$ & Dual background $\wh F_\xi
(s,\ghat)$ & Comments\\
\hline \hline $S_{2,1}$ & $\begin{array}{c} \frac{1}{2} (1+s_1),
\,0,\, \sqrt{s_2}, \,\frac{1}{2} (-1+s_1),\\ s_1 \neq 0,\, s_2> 0 \end{array}$  &
$\left(
\begin{array}{cccc}
 -\frac{1}{4 s_1} & 0 & 0 & \frac{1}{2 s_1} \\
 0 & \frac{1}{4 s_2} & 0 & 0 \\
 0 & 0 & \frac{1}{s_2} & 0 \\
 -\frac{1}{2 s_1} & 0 & 0 & \frac{1}{s_1}
\end{array}
\right)$ &   {\prop{diagonal metric}{torsionless}}\\
\hline $S_{2,2}$ & $\begin{array}{c} -\frac{s_2}{2 s_1}, \,0, \,0,\, s_1+\frac{s_2}{2 s_1},\\ s_1 \neq 0\end{array}$ & $\left(
\begin{array}{cccc}
 1-\frac{s_2}{s_1^2} & \frac{1}{2 s_1} & 0 & 0 \\
 \frac{1}{2 s_1} & 0 & 0 & 0 \\
 0 & 0 & \frac{1}{s_1^2} & 0 \\
 0 & 0 & 0 & \frac{1}{s_1^2}
\end{array}
\right)$ & $\begin{array}{c} \text{pp-wave metric}, \\ \text{torsionless} \end{array}$ \\
\hline
$S_{2,3}$ &  & & not dualizable \\
\hline
$S_{2,4}$ & $\begin{array}{c} s_1, \,0, \, \sqrt{s_2}, \, 0,\\ s_2>0\end{array}$ & $\text{diag}\left(-1,\frac{1}{4 s_2},\frac{1}{s_2},1\right)$ & {\prop{diagonal metric}{torsionless}}\\
\hline
$S_{2,5}$ & $\begin{array}{c} 0, \,0, \,\sqrt{s_2}, \,s_1,\\ s_2>0\end{array}$ & $\text{diag}\left(1,\frac{1}{4 s_2},\frac{1}{s_2},-1\right)$ & {\prop{diagonal metric}{torsionless}}\\
\hline $S_{2,6}$ & $\begin{array}{c} \frac{1}{2} (1+s_1), \,0, s_2, \,\frac{1}{2} (-1+s_1),\\ s_1\neq 0\end{array}$ & $\left(
\begin{array}{cccc}
 -\frac{1}{4 s_1} & 0 & \frac{1}{2 s_1} & 0 \\
 0 & 1 & 0 & 0 \\
 -\frac{1}{2 s_1} & 0 & \frac{1}{s_1} & 0 \\
 0 & 0 & 0 & 1
\end{array}
\right)$ &   {\prop{diagonal metric}{torsionless}}\\
\hline
$S_{2,7}$ &  & & not dualizable\\
\hline $S_{2,8}$ & $\begin{array}{c} -\frac{s_2}{2 s_1}, \,0, \,0,\, s_1+\frac{s_2}{2 s_1},\\ s_1 \neq 0 \end{array}$ & $\left(
\begin{array}{cccc}
 1-\frac{s_2}{s_1^2} & \frac{1}{2 s_1} & 0 & 0 \\
 \frac{1}{2 s_1} & 0 & 0 & 0 \\
 0 & 0 & \frac{1}{s_1^2} & 0 \\
 0 & 0 & 0 & 1
\end{array}
\right)$ & $\begin{array}{c} \text{pp-wave metric}, \\ \text{torsionless} \end{array}$ \\
\hline
$S_{2,9}$ &  & & not dualizable \\
\hline
$S_{2,10}$ & $0, \,s_1, \,s_2, \,0$ & $\text{diag}\left(1,1,-1,1\right)$ & $\begin{array}{c} \text{flat metric}, \\ \text{torsionless} \end{array}$ \\
\hline
$S_{2,11}$ & $s_1, \,0, \,0, \,s_2$ & $\text{diag}\left(-1,1,1,1\right)$ & $\begin{array}{c} \text{flat metric}, \\ \text{torsionless} \end{array}$ \\
\hline $S_{2,12}$ & $\begin{array}{c} -\frac{s_2}{1+s_1}, \,0, \,0,\, s_1+\frac{s_2}{1+s_1},\\ s_1 \neq 0,\, s_1 \neq -1 \end{array}$ &$\left(
\begin{array}{cccc}
 1-\frac{2 s_2}{(1+s_1)^2} & \frac{1}{1+s_1} & 0 & 0 \\
 \frac{1}{1+s_1} & 0 & 0 & 0 \\
 0 & 0 & \frac{1}{s_1^2} & 0 \\
 0 & 0 & 0 & \frac{1}{(1+s_1)^2}
\end{array}
\right)$ & $\begin{array}{c} \text{pp-wave metric}, \\ \text{torsionless} \end{array}$ \\
\hline $S_{2,13}$ & $0, \,s_2, \,s_1, \,0$ & $\left(
\begin{array}{cccc}
 1 & 0 & 0 & -\frac{s_2}{2 \epsilon } \\
 0 & 1 & 0 & \frac{s_1}{2 \epsilon } \\
 0 & 0 & 0 & -\frac{1}{2 \epsilon } \\
 \frac{s_2}{2 \epsilon } & -\frac{s_1}{2 \epsilon } & -\frac{1}{2 \epsilon } & -\frac{s_1^2+s_2^2}{4 \epsilon ^2}
\end{array}
\right)$ & {\prop{pp-wave metric}{with torsion}} \\
\hline $S_{2,14}$ & $\begin{array}{c} 0, \,s_2, \,s_1, \,0, \\ s_1^2+s_2^2 \neq \alpha^2 \end{array}$ & $\left(
\begin{array}{cccc}
\frac{-\alpha ^2+s_1^2}{-\alpha ^2+s_1^2+s_2^2} & \frac{s_1 s_2}{-\alpha ^2+s_1^2+s_2^2} & \frac{s_2}{-\alpha ^2+s_1^2+s_2^2} & 0 \\
 \frac{s_1 s_2}{-\alpha ^2+s_1^2+s_2^2} & \frac{-\alpha ^2+s_2^2}{-\alpha ^2+s_1^2+s_2^2} & \frac{-s_1}{-\alpha ^2+s_1^2+s_2^2} & 0 \\
 \frac{-s_2}{-\alpha ^2+s_1^2+s_2^2} & \frac{s_1}{-\alpha ^2+s_1^2+s_2^2} & \frac{1}{-\alpha ^2+s_1^2+s_2^2} & 0 \\
 0 & 0 & 0 & 1
\end{array}
\right)$ & {\prop{diagonalizable metric}{with torsion}}\\
\hline
\end{tabular}\normalsize \caption{Dual backgrounds for two-dimensional isometry groups. Part 1\label{table5}}
}
\end{center}
\end{table}

\begin{table}
\scriptsize {\renewcommand{\arraystretch}{1.4}
\begin{center}
\begin{tabular}{|c || c | c | c |}
\hline
 & Choice of $\xi^{0}, \,\xi^{1}, \,\xi^{2}, \,\xi^{3}$ & Dual background $\wh F_\xi
(s,\ghat)$ & Comments \\
\hline \hline $S_{2,15}$ & $\begin{array}{c} 0, \,0, \,\sqrt{s_1}, \,-s_2, \\ s_1> 0\end{array}$
& $\left(
\begin{array}{cccc}
 \frac{1}{4 s_1} & 0 & 0 & 0 \\
 0 & \frac{s_1}{\alpha ^2+s_1} & \frac{\alpha }{\alpha ^2+s_1} & 0 \\
 0 & \frac{-\alpha }{\alpha ^2+s_1} & \frac{1}{\alpha ^2+s_1} & 0 \\
 0 & 0 & 0 & -1
\end{array}
\right)$ & {\prop{diagonalizable metric}{with torsion}}\\
\hline
$S_{2,16}$& $\begin{array}{c} \frac{1}{2} (1-s_1), \,0,\\  s_2, \,\frac{1}{2} (-1-s_1),\\ s_1 \neq \alpha^2 \end{array}$ & $\left(
\begin{array}{cccc}
 \frac{-1}{4 \left(\alpha ^2-s_1\right)} & \frac{-\alpha }{2 \left(\alpha ^2-s_1\right)} & \frac{-1}{2 \left(\alpha ^2-s_1\right)} & 0 \\
 \frac{-\alpha }{2 \left(\alpha ^2-s_1\right)} & \frac{-s_1}{\alpha ^2-s_1} & \frac{-\alpha }{\alpha ^2-s_1} & 0 \\
 \frac{1}{2 \alpha ^2-2 s_1} & \frac{\alpha }{\alpha ^2-s_1} & \frac{1}{\alpha ^2-s_1} & 0 \\
 0 & 0 & 0 & 1
\end{array}
\right)$ & {\prop{diagonalizable metric}{with torsion}}\\
\hline $S_{2,17}$ & $0, \,s_1, \,s_2, \,0$ & $\left(
\begin{array}{cccc}
 1 & 0 & 0 & 0 \\
 0 & 1 & 0 & 0 \\
 0 & 0 & 0 & -\frac{1}{2} \\
 0 & 0 & -\frac{1}{2} & -s_1
\end{array}
\right)$ & $\begin{array}{c} \text{flat metric}, \\ \text{torsionless} \end{array}$ \\
\hline
$S_{2,18}$ &  & & not dualizable \\
\hline $S_{2,19}$ & $\begin{array}{c} -\frac{s_2}{12}, \,\frac{s_1}{4}, \,0,
\,\frac{s_2}{12}, \\ s_1 \neq 0 \end{array}$ & $\left(
\begin{array}{cccc}
 \frac{1}{16} & 0 & 0 & 0 \\
 0 & -\frac{1}{36 s_1} & -\frac{1}{6 s_1} & 0 \\
 0 & \frac{1}{6 s_1} & \frac{1}{s_1} & 0 \\
 0 & 0 & 0 & 1
\end{array}
\right)$ & {\prop{diagonalizable metric}{with torsion}}\\
\hline $S_{2,20}$ & $\begin{array}{c}\frac{\sqrt{|s_1|}(1+\sigma)}{2}, 0,\\ s_2, \frac{\sqrt{|s_1|}(1-\sigma)}{2},\\ s_1\neq 0,\\ \sigma=\text{sgn}(s_1) \end{array}$ & $\left(
\begin{array}{cccc}
 -\frac{1}{4 s_1} & 0 & 0 & 0 \\
 0 & 1 & 0 & 0 \\
 0 & 0 & \frac{\sigma s_1}{\sigma s_1^2+\hat{g}_2^2} & \frac{-\hat{g}_2}{\sigma s_1^2+\hat{g}_2^2} \\
 0 & 0 & \frac{\hat{g}_2}{\sigma s_1^2+\hat{g}_2^2} & \frac{s_1}{\sigma s_1^2+\hat{g}_2^2}
\end{array}
\right)$ & {\prop{diagonalizable metric}{with torsion}}\\
\hline $S_{2,21}$ & $\frac{1}{2}, \,s_2, \,-s_1, \,\frac{1}{2}$ &
$\left(
\begin{array}{cccc}
 1 & 0 & 0 & \frac{\cot(\gamma) s_2}{-1+\hat{g}_2} \\
 0 & 1 & 0 & \frac{\cot(\gamma) s_1}{1-\hat{g}_2} \\
 0 & 0 & 0 & \frac{\csc(\gamma)}{1-\hat{g}_2} \\
 \frac{\cot(\gamma) s_2}{1+\hat{g}_2} & -\frac{\cot(\gamma) s_1}{1+\hat{g}_2} & \frac{\csc(\gamma)}{1+\hat{g}_2} & \frac{\cot(\gamma)^2 \left(s_1^2+s_2^2\right)}{-1+\hat{g}_2^2}
\end{array}
\right)$ &  {\prop{pp-wave metric}{with torsion}} \\
\hline $S_{2,22}$ & $\frac{1}{2}, \,s_1, \,s_2, \,\frac{1}{2}$ &
$\left(
\begin{array}{cccc}
 1 & 0 & 0 & 0 \\
 0 & 1 & 0 & 0 \\
 0 & 0 & 0 & \frac{1}{1-\hat{g}_2} \\
 0 & 0 & \frac{1}{1+\hat{g}_2} & 0
\end{array}
\right)$ & $\begin{array}{c} \text{flat metric}, \\ \text{torsionless} \end{array}$ \\
\hline $S_{2,23}$ & $\begin{array}{c} \frac{1}{2} (1-s_1), \,0,\\
s_2, \,\frac{1}{2} (1+s_1), \\ s_1 \neq \alpha^2 \end{array}$ &
$\left(
\begin{array}{cccc}
 \frac{-1}{4 \left(\alpha ^2-s_1+\hat{g}_2^2\right)} & \frac{\alpha }{2 \left(\alpha ^2-s_1+\hat{g}_2^2\right)} & \frac{1}{2 \alpha ^2-2 s_1+2 \hat{g}_2^2} & \frac{-\hat{g}_2}{2 \left(\alpha ^2-s_1+\hat{g}_2^2\right)} \\
 \frac{\alpha }{2 \left(\alpha ^2-s_1+\hat{g}_2^2\right)} & \frac{-s_1+\hat{g}_2^2}{\alpha ^2-s_1+\hat{g}_2^2} & \frac{-\alpha }{\alpha ^2-s_1+\hat{g}_2^2} & \frac{\alpha  \hat{g}_2}{\alpha ^2-s_1+\hat{g}_2^2} \\
 \frac{-1}{2 \left(\alpha ^2-s_1+\hat{g}_2^2\right)} & \frac{\alpha }{\alpha ^2-s_1+\hat{g}_2^2} & \frac{1}{\alpha ^2-s_1+\hat{g}_2^2} & \frac{-\hat{g}_2}{\alpha ^2-s_1+\hat{g}_2^2} \\
 \frac{-\hat{g}_2}{2 \left(\alpha ^2-s_1+\hat{g}_2^2\right)} & \frac{\alpha  \hat{g}_2}{\alpha ^2-s_1+\hat{g}_2^2} & \frac{\hat{g}_2}{\alpha ^2-s_1+\hat{g}_2^2} & \frac{\alpha ^2-s_1}{\alpha ^2-s_1+\hat{g}_2^2}
\end{array}
\right)$ & {\propp{in principle}{diagonalizable metric}{with torsion}}\\

%\hline $S_{2,23}$ & $\begin{array}{c} \frac{\alpha^2-2s_1}{2\sqrt{|\alpha^2-s_1|}}, \ 0,\\
%s_2-\log\sqrt{|\alpha^2-s_1|}, \\
%\frac{\alpha^2}{2\sqrt{|\alpha^2-s_1|}}, \\ s_1 \neq \alpha^2,\\
%\sigma=\text{sgn}(\alpha^2-s_1)
%\end{array}$ & $\left(
%\begin{array}{cccc}
% -\frac{1}{4 \alpha^2-4 {s_1}} & \frac{a}{2 \alpha^2-2 {s_1}} & 0 & 0 \\
% \frac{a}{2 \alpha^2-2 {s_1}} & \frac{{\hat{g}_2}^2+{s_1} \left({s_1}-\alpha^2\right) \sigma
%   }{{\hat{g}_2}^2+\left(\alpha^2-{s_1}\right)^2 \sigma } & \frac{a \left({s_1}-\alpha^2\right) \sigma
%   }{{\hat{g}_2}^2+\left(\alpha^2-{s_1}\right)^2 \sigma } & \frac{a {\hat{g}_2}}{{\hat{g}_2}^2+\left(\alpha^2-{s_1}\right)^2
%   \sigma } \\
% 0 & \frac{a \left(\alpha^2-{s_1}\right) \sigma }{{\hat{g}_2}^2+\left(\alpha^2-{s_1}\right)^2 \sigma } &
%   \frac{\left(\alpha^2-{s_1}\right) \sigma }{{\hat{g}_2}^2+\left(\alpha^2-{s_1}\right)^2 \sigma } &
%   -\frac{{\hat{g}_2}}{{\hat{g}_2}^2+\left(\alpha^2-{s_1}\right)^2 \sigma } \\
% 0 & \frac{a {\hat{g}_2}}{{\hat{g}_2}^2+\left(\alpha^2-{s_1}\right)^2 \sigma } &
%   \frac{{\hat{g}_2}}{{\hat{g}_2}^2+\left(\alpha^2-{s_1}\right)^2 \sigma } &
%   \frac{\alpha^2-{s_1}}{{\hat{g}_2}^2+\left(\alpha^2-{s_1}\right)^2 \sigma }
%\end{array}
%\right)$ & {\propp{in principle}{diagonalizable metric}{with torsion}}\\

\hline $S_{2,24}$ & $\begin{array}{c} s_2, \,s_1, \,0, \,0, \\
s_2\neq 0 \end{array}$ & $\left(
\begin{array}{cccc}
 1 & 0 & 0 & 0 \\
 0 & \frac{\alpha ^2+\hat{g}_2^2}{s_2^2-\hat{g}_2^2} & \frac{1}{s_2+\hat{g}_2} & -\frac{\alpha ^2+s_2^2}{s_2^2-\hat{g}_2^2} \\
 0 & \frac{1}{-s_2+\hat{g}_2} & 0 & \frac{1}{s_2-\hat{g}_2} \\
 0 & \frac{\alpha ^2+s_2^2}{s_2^2-\hat{g}_2^2} & \frac{1}{s_2+\hat{g}_2} & -\frac{\alpha ^2+s_2^2}{s_2^2-\hat{g}_2^2}
\end{array}
\right)$ & $\begin{array}{c} \text{flat metric}, \\ \text{torsionless} \end{array}$\\
\hline
\end{tabular}
\normalsize \caption{Dual backgrounds for two-dimensional isometry
groups. Part 2\label{table6}}
\end{center}
}
\end{table}

\newcommand{\mm}[2]{\begin{array}{c}#1,\\#2\\ \end{array}}
\newcommand{\mmm}[3]{\begin{array}{c}#1,\\#2,\\#3\\ \end{array}}

\begin{table}
\begin{center}
\scriptsize {\renewcommand{\arraystretch}{1.4}
\begin{tabular}{|c||c|c|c|}
\hline
%Sub-group
& Choice of $\xi^{0}, \xi^{1}, \xi^{2}, \xi^{3}$ & Dual background
$\wh F_\xi
(s,\ghat)$ & Comments \\
\hline \hline
$S_{3,1}$ & & &  not dualizable   \\
\hline
$S_{3,2}$ & $\mm{0, \sqrt{s_1}, 0, 0}{s_1>0}$ & $\text{diag}\left(\frac{1}{4s_1},\frac{1}{s_1},-1,1\right)$ & \prop{diagonal metric}{torsionless} \\
\hline $S_{3,3}$ &$\mm{\frac{1+s_1}{2}, 0, 0, \frac{1-s_1}{2}}{s_1\neq 0}$ &
$\left(
\begin{array}{cccc}
 -\frac{1}{4 s_1} & -\frac{1}{2 s_1} & 0 & 0 \\
 \frac{1}{2 s_1} & \frac{1}{s_1} & 0 & 0 \\
 0 & 0 & 1 & 0 \\
 0 & 0 & 0 & 1 \\
\end{array}
\right)$ & \prop{diagonal metric}{torsionless}  \\
\hline
$S_{3,4}$ & & & not dualizable \\
\hline
$S_{3,5}$ & & & not dualizable \\
\hline
$S_{3,6}$ & $s_1,0,0,0$ & \diag{-1}{1}{1}{1} & \prop{flat metric}{torsionless} \\
\hline
$S_{3,7}$ & $0,0,0,s_1$ & \diag{1}{-1}{1}{1} & \prop{flat metric}{torsionless} \\
\hline
$S_{3,8}$ & & & not dualizable \\
\hline $S_{3,9}$ & $0,-\frac{s_1}{2},0,0$ & $\left(
\begin{array}{cccc}
 \frac{1}{4} & 0 & 0 & 0 \\
 0 & 0 & 1 & 0 \\
 0 & 1 & -s_1 & 0 \\
 0 & 0 & 0 & 1 \\
\end{array}
\right)$ & \prop{flat metric}{torsionless} \\
\hline $S_{3,10}$ & $\mmm{\frac{\sqrt{|s_1|}(1+\sigma)}{2}, 0, 0,
\frac{\sqrt{|s_1|}(1-\sigma)}{2}}{s_1\neq 0}{\sigma=\text{sgn}(s_1)}$ &$\left(
\begin{array}{cccc}
 -\frac{1}{4 s_1} & 0 & 0 & 0 \\
 0 & \frac{\sigma s_1}{\hat{g}_2^2+\sigma s_1^2 } &
   -\frac{\hat{g}_2}{\hat{g}_2^2+\sigma s_1^2} & 0 \\
 0 & \frac{\hat{g}_2}{\hat{g}_2^2+\sigma s_1^2} &
   \frac{s_1}{\hat{g}_2^2+\sigma s_1^2} & 0 \\
 0 & 0 & 0 & 1 \\
\end{array}
\right)$ & \prop{diagonal metric}{with torsion}\\
\hline $S_{3,11}$ &$\mm{\frac{1}{2}, \sqrt{s_1}, 0, \frac{1}{2}}{s_1>0}$
&$\left(
\begin{array}{cccc}
 \frac{1}{4 s_1} & 0 & 0 & 0 \\
 0 & 0 & \frac{1}{1-\hat{g}_2} & 0 \\
 0 & \frac{1}{1+\hat{g}_2} & 0 & 0 \\
 0 & 0 & 0 & \frac{1}{s_1} \\
\end{array}
\right)$ & \prop{diagonalizable metric}{torsionless} \\
\hline $S_{3,12}$ &$\frac{1}{2}, s_1, 0, \frac{1}{2}$ &$\left(
\begin{array}{cccc}
 1 & 0 & 0 & 0 \\
 0 & 0 & \frac{1}{1-\hat{g}_2} & 0 \\
 0 & \frac{1}{1+\hat{g}_2} & 0 & 0 \\
 0 & 0 & 0 & 1 \\
\end{array}
\right)$ & \prop{flat metric}{torsionless}\\
\hline $S_{3,13}$ &$\mm{s_1,0,0,0}{s_1\neq 0}$ &$\left(
\begin{array}{cccc}
 \frac{\alpha ^2+\hat{g}_2^2}{s_1^2-\hat{g}_2^2} & \frac{1}{s_1+\hat{g}_2} & -\frac{\alpha ^2+s_1^2}{s_1^2-\hat{g}_2^2} & 0 \\
 \frac{1}{\hat{g}_2-s_1} & 0 & \frac{1}{s_1-\hat{g}_2} & 0 \\
 \frac{\alpha ^2+s_1^2}{s_1^2-\hat{g}_2^2} & \frac{1}{s_1+\hat{g}_2} & -\frac{\alpha ^2+s_1^2}{s_1^2-\hat{g}_2^2} & 0 \\
 0 & 0 & 0 & 1 \\
\end{array}
\right)$ & \prop{flat metric}{torsionless} \\
\hline
$S_{3,14}$ & & & not transitive action \\
\hline
$S_{3,15}$ & & & not dualizable \\
\hline
$S_{3,16}$ & & & not dualizable \\
\hline
$S_{3,17}$ & & & not dualizable \\
\hline $S_{3,18}$ & $0,0,s_1,0$ &$\left(
\begin{array}{cccc}
 1 & 0 & 0 & 0 \\
 0 & 0 & 0 & 1 \\
 0 & 0 & 1 & -\hat{g}_3 \\
 0 & 1 & \hat{g}_3 & -\hat{g}_3^2 \\
\end{array}
\right)$ & \prop{flat metric}{torsionless} \\
\hline
$S_{3,19}$ & & & not dualizable \\
\hline
\end{tabular}
\caption{Dual backgrounds for three-dimensional isometry groups.
Part 1\label{table7} } }\end{center}
\end{table}

\begin{table}
\begin{center}
\scriptsize {\renewcommand{\arraystretch}{1.4}
\begin{tabular}{|c || c | c | c |}
\hline
%Sub-group
& Choice of $\xi^{0}, \xi^{1}, \xi^{2}, \xi^{3}$ & Dual background
$\wh F_\xi
(s,\ghat)$ & Comments \\
\hline \hline $S_{3,20}$ &$0,-\frac{s_1}{2},0,0$ & $\left(
\begin{array}{cccc}
 \frac{1}{4 \beta ^2+4} & 0 & -\frac{\beta }{2 \beta ^2+2} & -\frac{\beta ^2 \hat{g}_3}{2 \beta ^2+2} \\
 0 & 0 & 0 & 1 \\
 \frac{\beta }{2 \beta ^2+2} & 0 & \frac{1}{\beta ^2+1} & \frac{\beta  \hat{g}_3}{\beta ^2+1} \\
 -\frac{\beta ^2 \hat{g}_3}{2 \beta ^2+2} & 1 & -\frac{\beta  \hat{g}_3}{\beta ^2+1} & -\frac{\beta ^2 \hat{g}_3^2}{\beta ^2+1}-s_1 \\
\end{array}
\right)$ & \prop{flat metric}{torsionless} \\
\hline
$S_{3,21}$ & & & not dualizable \\
\hline $S_{3,22}$ & $\frac{1}{2},0,s_1,\frac{1}{2}$ & $\left(
\begin{array}{cccc}
 1 & 0 & 0 & 0 \\
 0 & 0 & 0 & \frac{1}{1-\hat{g}_3} \\
 0 & 0 & 1 & \frac{\hat{g}_3 \alpha +\alpha +\hat{g}_2}{1-\hat{g}_3} \\
 0 & \frac{1}{\hat{g}_3+1} & \frac{-\hat{g}_3 \alpha +\alpha -\hat{g}_2}{\hat{g}_3+1} & \frac{\left(\hat{g}_2+\alpha  \hat{g}_3\right){}^2}{\hat{g}_3^2-1} \\
\end{array}
\right)$ & \prop{pp-wave metric}{torsionless} \\
\hline $S_{3,23}$ &$\mmm{\frac{\sigma}{2},0,-\alpha \log |s_1|,\frac{\sigma}{2}}{s_1\neq 0}{\sigma=\text{sgn}(s_1)}$ &
$\left(
\begin{array}{cccc}
 \frac{\alpha ^2}{s_1^2} & 0 & 0 & -\frac{\alpha ^2}{s_1 \left(\sigma-\hat{g}_3\right)} \\
 0 & 0 & 0 & \frac{1}{\sigma-\hat{g}_3} \\
 0 & 0 & 1 & \frac{\beta  \sigma+\hat{g}_2+\beta  \hat{g}_3}{\sigma-\hat{g}_3} \\
 \frac{\alpha ^2}{s_1 \left(\sigma+\hat{g}_3\right)} & \frac{1}{\sigma+\hat{g}_3} & -\frac{-\beta
   \sigma+\hat{g}_2+\beta  \hat{g}_3}{\sigma+\hat{g}_3} & -\frac{\alpha ^2+\left(\hat{g}_2+\beta \hat{g}_3\right)^2}{1-\hat{g}_3^2} \\
\end{array}
\right)$ & \prop{pp-wave metric}{torsionless} \\
\hline $S_{3,24}$ & $\mm{0,0,0,\sqrt{|s_1|}}{s_1\neq
0}$ & $\left(
\begin{array}{cccc}
 \frac{1}{4 {s_1}} & 0 & 0 & 0 \\
 0 & \frac{|{s_1} |
   }{{\hat{g}_2}^2+{\hat{g}_3}^2-{s_1}|s_1|  } &
   -\frac{{\hat{g}_2}}{{\hat{g}_2}^2+{\hat{g}_3}^2-{s_1}|s_1|} &
   -\frac{{\hat{g}_3}}{{\hat{g}_2}^2+{\hat{g}_3}^2-{s_1}|s_1| } \\
 0 &
   \frac{{\hat{g}_2}}{{\hat{g}_2}^2+{\hat{g}_3}^2-{s_1}|s_1|
   } & \frac{{\hat{g}_3}^2-{s_1}|s_1|}{({\hat{g}_2}^2+{\hat{g}_3}^2) |  {s_1}|- {s_1}^3} &
   \frac{-{\hat{g}_2}\hat{g}_3}{({\hat{g}_2}^2+{\hat{g}_3}^2) |  {s_1}|- {s_1}^3} \\
 0 &
   \frac{{\hat{g}_3}}{{\hat{g}_2}^2+{\hat{g}_3}^2-{s_1}|s_1|
    } & \frac{-{\hat{g}_2} {\hat{g}_3}}{({\hat{g}_2}^2+{\hat{g}_3}^2) |  {s_1}|- {s_1}^3} & \frac{{\hat{g}_2}^2-{s_1}|s_1| }{({\hat{g}_2}^2+{\hat{g}_3}^2) |  {s_1}|- {s_1}^3}
\end{array}
\right)$ & {\propp{in principle}{diagonalizable metric}{with torsion}} \\
\hline $S_{3,25}$ & $\frac{1}{2},0,s_1,\frac{1}{2}$ & $\left(
\begin{array}{cccc}
 1 & 0 & 0 & 0 \\
 0 & 0 & 0 & \frac{1}{1-\hat{g}_3} \\
 0 & 0 & 1 & \frac{\hat{g}_2}{1-\hat{g}_3} \\
 0 & \frac{1}{\hat{g}_3+1} & -\frac{\hat{g}_2}{\hat{g}_3+1} & \frac{\hat{g}_2^2}{\hat{g}_3^2-1} \\
\end{array}
\right)$ & \prop{pp-wave metric}{torsionless} \\
\hline $S_{3,26}$ & $\mmm{\frac{\sigma}{2},0,\alpha \log |s_1|,\frac{\sigma}{2}}{s_1\neq 0}{\sigma=\text{sgn}(s_1)}$ &
$\left(
\begin{array}{cccc}
 \frac{\alpha ^2}{s_1^2} & 0 & 0 & -\frac{\alpha ^2}{s_1 \left(\sigma-\hat{g}_3\right)} \\
 0 & 0 & 0 & \frac{1}{\sigma-\hat{g}_3} \\
 0 & 0 & 1 & \frac{\hat{g}_2}{\sigma-\hat{g}_3} \\
 \frac{\alpha ^2}{s_1 \left(\sigma+\hat{g}_3\right)} & \frac{1}{\sigma+\hat{g}_3} &
   -\frac{\hat{g}_2}{\sigma+\hat{g}_3} & -\frac{\alpha ^2+\hat{g}_2^2}{1-\hat{g}_3^2} \\
\end{array}
\right)$ & \prop{pp-wave metric}{torsionless} \\
\hline $S_{3,27}$ & $\mm{0,\sqrt{s_1},0,0}{s_1>0}$ & see $\wh F_{3,27}
(s,\ghat)$ below  & \prop{diagonalizable
%$2\times2$
metric}{with torsion} \\
\hline
$S_{3,28}$ & & & not transitive action \\
\hline $S_{3,29}$ & $0,s_1,0,0$ & $\left(
\begin{array}{cccc}
 1 & 0 & 0 & 0 \\
 0 & \frac{1}{\alpha ^2+\hat{g}_2^2-\hat{g}_3^2} & -\frac{\hat{g}_3}{\alpha ^2+\hat{g}_2^2-\hat{g}_3^2} & \frac{\hat{g}_2}{\alpha ^2+\hat{g}_2^2-\hat{g}_3^2} \\
 0 & \frac{\hat{g}_3}{\alpha ^2+\hat{g}_2^2-\hat{g}_3^2} & -\frac{\alpha ^2+\hat{g}_2^2}{\alpha ^2+\hat{g}_2^2-\hat{g}_3^2} & \frac{\hat{g}_2 \hat{g}_3}{\alpha
   ^2+\hat{g}_2^2-\hat{g}_3^2} \\
 0 & -\frac{\hat{g}_2}{\alpha ^2+\hat{g}_2^2-\hat{g}_3^2} & \frac{\hat{g}_2 \hat{g}_3}{\alpha ^2+\hat{g}_2^2-\hat{g}_3^2} & \frac{\alpha ^2-\hat{g}_3^2}{\alpha
   ^2+\hat{g}_2^2-\hat{g}_3^2} \\
\end{array}
\right)$ & \prop{diagonalizable
%$2\times2$
metric}{with torsion} \\
\hline
$S_{3,30}$ & & & not transitive action \\
\hline $S_{3,31}$ & $\mmm{\frac{\sqrt{|s_1|}(1+\sigma)}{2}, 0, 0,
\frac{\sqrt{|s_1|}(1-\sigma)}{2}}{s_1\neq
0}{\sigma=\text{sgn}(s_1)}$ & see $\wh F_{3,31}
(s,\ghat)$ below & {\propp{in principle}{diagonalizable metric}{with torsion}} \\
\hline
\end{tabular}
\normalsize \caption{Dual backgrounds for three-dimensional isometry
groups. Part 2\label{table8}} }\end{center}
\end{table}

\begin{table}
\begin{center}
\scriptsize {\renewcommand{\arraystretch}{1.4}
\begin{tabular}{|c || c | c | c |}
\hline
%Sub-group
& Choice of $\xi^{0}, \xi^{1}, \xi^{2}, \xi^{3}$ & Dual background
$\wh F_\xi
(s,\ghat)$ & Comments \\
\hline \hline
$S_{3,32}$ & & & not transitive action \\
\hline
$S_{3,33}$ & & & not dualizable \\
\hline
$S_{3,34}$ & & & not dualizable \\
\hline $S_{3,35}$ & $0,0,0,s_1$ & $\left(
\begin{array}{cccc}
 1 & 0 & 0 & 0 \\
 0 & \frac{1}{-\alpha ^2+\hat{g}_2^2+\hat{g}_3^2} & \frac{\hat{g}_3}{-\alpha ^2+\hat{g}_2^2+\hat{g}_3^2} & -\frac{\hat{g}_2}{-\alpha ^2+\hat{g}_2^2+\hat{g}_3^2} \\
 0 & -\frac{\hat{g}_3}{-\alpha ^2+\hat{g}_2^2+\hat{g}_3^2} & \frac{\hat{g}_2^2-\alpha ^2}{-\alpha ^2+\hat{g}_2^2+\hat{g}_3^2} & \frac{\hat{g}_2 \hat{g}_3}{-\alpha
   ^2+\hat{g}_2^2+\hat{g}_3^2} \\
 0 & \frac{\hat{g}_2}{-\alpha ^2+\hat{g}_2^2+\hat{g}_3^2} & \frac{\hat{g}_2 \hat{g}_3}{-\alpha ^2+\hat{g}_2^2+\hat{g}_3^2} & \frac{\hat{g}_3^2-\alpha ^2}{-\alpha
   ^2+\hat{g}_2^2+\hat{g}_3^2} \\
\end{array}
\right)$ & \prop{diagonalizable
%$2\times2$
metric}{with torsion} \\
\hline $S_{3,36}$ & $s_1,0,0,0$ & $\left(
\begin{array}{cccc}
 -1 & 0 & 0 & 0 \\
 0 & \frac{1}{\alpha ^2+\hat{g}_2^2+\hat{g}_3^2} & \frac{\hat{g}_3}{\alpha ^2+\hat{g}_2^2+\hat{g}_3^2} & -\frac{\hat{g}_2}{\alpha ^2+\hat{g}_2^2+\hat{g}_3^2} \\
 0 & -\frac{\hat{g}_3}{\alpha ^2+\hat{g}_2^2+\hat{g}_3^2} & \frac{\alpha ^2+\hat{g}_2^2}{\alpha ^2+\hat{g}_2^2+\hat{g}_3^2} & \frac{\hat{g}_2 \hat{g}_3}{\alpha
   ^2+\hat{g}_2^2+\hat{g}_3^2} \\
 0 & \frac{\hat{g}_2}{\alpha ^2+\hat{g}_2^2+\hat{g}_3^2} & \frac{\hat{g}_2 \hat{g}_3}{\alpha ^2+\hat{g}_2^2+\hat{g}_3^2} & \frac{\alpha ^2+\hat{g}_3^2}{\alpha
   ^2+\hat{g}_2^2+\hat{g}_3^2} \\
\end{array}
\right)$ & \prop{diagonalizable
%$2\times2$
metric}{with torsion} \\
\hline $S_{3,37}$ & $\mmm{\frac{\sqrt{|s_1|}(1+\sigma)}{2}, 0, 0,
\frac{\sqrt{|s_1|}(1-\sigma)}{2}}{s_1\neq
0}{\sigma=\text{sgn}(s_1)}$   & see $\wh F_{3,37}
(s,\ghat)$ below  & {\propp{in principle}{diagonalizable metric}{with torsion}} \\
\hline
$S_{3,38}$ & & & not transitive action \\
 \hline
$S_{3,39}$ & & & not transitive action \\
\hline
\end{tabular}
\normalsize \caption{Dual backgrounds for three-dimensional isometry
groups. Part 3 \label{table9}} }\end{center}
\end{table}

\newpage
{\footnotesize $$\wh F_{3,27} (s,\ghat)=\left(
\begin{array}{cccc}
 \frac{1}{4 s_1} & 0 & 0 & 0 \\
 0 & \frac{1}{s_1 \cos ^2(\gamma )+\sin ^2(\gamma ) \left(\hat{g}_2^2-\hat{g}_3^2\right)} & \frac{\sin (\gamma ) \hat{g}_3}{\sin ^2(\gamma )
   \left(\hat{g}_3^2-\hat{g}_2^2\right)-\cos ^2(\gamma ) s_1} & \frac{\sin (\gamma ) \hat{g}_2}{s_1 \cos ^2(\gamma )+\sin ^2(\gamma ) \left(\hat{g}_2^2-\hat{g}_3^2\right)}
   \\
 0 & -\frac{\sin (\gamma ) \hat{g}_3}{\sin ^2(\gamma ) \left(\hat{g}_3^2-\hat{g}_2^2\right)-\cos ^2(\gamma ) s_1} & -\frac{s_1 \cos ^2(\gamma )+\sin ^2(\gamma )
   \hat{g}_2^2}{s_1 \cos ^2(\gamma )+\sin ^2(\gamma ) \left(\hat{g}_2^2-\hat{g}_3^2\right)} & \frac{\sin ^2(\gamma ) \hat{g}_2 \hat{g}_3}{s_1 \cos ^2(\gamma )+\sin ^2(\gamma
   ) \left(\hat{g}_2^2-\hat{g}_3^2\right)} \\
 0 & -\frac{\sin (\gamma ) \hat{g}_2}{s_1 \cos ^2(\gamma )+\sin ^2(\gamma ) \left(\hat{g}_2^2-\hat{g}_3^2\right)} & \frac{\sin ^2(\gamma ) \hat{g}_2 \hat{g}_3}{s_1 \cos
   ^2(\gamma )+\sin ^2(\gamma ) \left(\hat{g}_2^2-\hat{g}_3^2\right)} & \frac{\cos ^2(\gamma ) s_1-\sin ^2(\gamma ) \hat{g}_3^2}{s_1 \cos ^2(\gamma )+\sin ^2(\gamma )
   \left(\hat{g}_2^2-\hat{g}_3^2\right)} \\
\end{array}
\right)$$}

{\footnotesize $$\wh F_{3,31} (s,\ghat)=\left(
\begin{array}{cccc}
 -\frac{1}{4 s_1} & 0 & 0 & 0 \\
 0 & \frac{1}{\left({\hat{g}_2}^2+{\hat{g}_3}^2\right) \cos ^2(\gamma )+s_1 \sin ^2(\gamma )} & \frac{{\hat{g}_3} \cos
   (\gamma )}{\left({\hat{g}_2}^2+{\hat{g}_3}^2\right) \cos ^2(\gamma )+s_1 \sin ^2(\gamma )} & -\frac{{\hat{g}_2} \cos
   (\gamma )}{\left({\hat{g}_2}^2+{\hat{g}_3}^2\right) \cos ^2(\gamma )+s_1 \sin ^2(\gamma )} \\
 0 & -\frac{{\hat{g}_3} \cos (\gamma )}{\left({\hat{g}_2}^2+{\hat{g}_3}^2\right) \cos ^2(\gamma )+s_1 \sin ^2(\gamma )}
   & \frac{{\hat{g}_2}^2 \cos ^2(\gamma )+s_1 \sin ^2(\gamma )}{\left({\hat{g}_2}^2+{\hat{g}_3}^2\right) \cos ^2(\gamma
   )+s_1 \sin ^2(\gamma )} & \frac{{\hat{g}_2} {\hat{g}_3} \cos ^2(\gamma )}{\left({\hat{g}_2}^2+{\hat{g}_3}^2\right)
   \cos ^2(\gamma )+s_1 \sin ^2(\gamma )} \\
 0 & \frac{{\hat{g}_2} \cos (\gamma )}{\left({\hat{g}_2}^2+{\hat{g}_3}^2\right) \cos ^2(\gamma )+s_1 \sin ^2(\gamma )}
   & \frac{{\hat{g}_2} {\hat{g}_3} \cos ^2(\gamma )}{\left({\hat{g}_2}^2+{\hat{g}_3}^2\right) \cos ^2(\gamma )+s_1 \sin
   ^2(\gamma )} & \frac{{\hat{g}_3}^2 \cos ^2(\gamma )+s_1 \sin ^2(\gamma )}{\left({\hat{g}_2}^2+{\hat{g}_3}^2\right)
   \cos ^2(\gamma )+s_1 \sin ^2(\gamma )} \\
\end{array}
\right)$$}

{\footnotesize $$\wh F_{3,37} (s,\ghat)=\left(
\begin{array}{cccc}
 -\frac{1}{4 s_1} & 0 & 0 & 0 \\
 0 & \frac{|s_1|}{s_1|s_1| \sin ^2(c)+{\hat{g}_2}^2+{\hat{g}_3}^2} & -\frac{{\hat{g}_2} \sin
   (\gamma)+{\hat{g}_3} \cos (\gamma)}{s_1|s_1| \sin ^2(\gamma)+{\hat{g}_2}^2+{\hat{g}_3}^2} & \frac{{\hat{g}_2} \cos
   (\gamma)-{\hat{g}_3} \sin (\gamma)}{s_1|s_1| \sin ^2(\gamma)+{\hat{g}_2}^2+{\hat{g}_3}^2} \\
 0 & \frac{{\hat{g}_2} \sin (\gamma)+{\hat{g}_3} \cos (\gamma)}{s_1|s_1| \sin ^2(\gamma)+{\hat{g}_2}^2+{\hat{g}_3}^2} &
   \frac{(\hat{g}_2\cos(\gamma)-\hat{g}_3 \sin (\gamma))^2 +s_1
   |s_1|\sin ^2(\gamma)}{|s_1| \left(s_1|s_1| \sin ^2(\gamma)+{\hat{g}_2}^2+{\hat{g}_3}^2\right)} &
   \frac{\sin (2\gamma) \left({\hat{g}_2}^2-{\hat{g}_3}^2\right)+2 {\hat{g}_2} {\hat{g}_3} \cos (2\gamma)}{2 |s_1|
   \left(s_1|s_1| \sin ^2(\gamma)+{\hat{g}_2}^2+{\hat{g}_3}^2\right)} \\
 0 & \frac{{\hat{g}_3} \sin (\gamma)-{\hat{g}_2} \cos (\gamma)}{s_1|s_1| \sin ^2(\gamma)+{\hat{g}_2}^2+{\hat{g}_3}^2} &
   \frac{\sin (2\gamma) \left({\hat{g}_2}^2-{\hat{g}_3}^2\right)+2 {\hat{g}_2} {\hat{g}_3} \cos (2\gamma)}{|s_1|
   \left(s_1|s_1| \sin ^2(\gamma)+{\hat{g}_2}^2+{\hat{g}_3}^2\right)} & \frac{(\hat{g}_2\cos(\gamma)+\hat{g}_3 \sin (\gamma))^2 +s_1
   |s_1|\sin ^2(\gamma)}{|s_1| \left(
   s_1|s_1| \sin ^2(\gamma)+{\hat{g}_2}^2+{\hat{g}_3}^2\right)} \\
\end{array}
\right)$$}

%%%%%%%%%%%%%%%%%%%%%%%%%%%%%%%%%%%%%%%%%%%%%%%%%%%%%%%%%%%%%%%%%%%%%%%%%%%%%%%
%% Backmatter
%%%%%%%%%%%%%%%%%%%%%%%%%%%%%%%%%%%%%%%%%%%%%%%%%%%%%%%%%%%%%%%%%%%%%%%%%%%%%%%

\end{document}